%% file: paper.tex
\documentclass[]{emulateapj}

\usepackage{natbib}
\usepackage{amsmath}
\usepackage{threeparttable}
\usepackage{amssymb}
\usepackage{graphicx}
\usepackage{longtable}
\usepackage{tabularx}

\newcommand{\HI}{\ensuremath{\mbox{\rm \ion{H}{1}}}}

\newcommand{\htwo}{\ensuremath{\mbox{H$_2$}}}
\newcommand{\msun}{\ensuremath{M_\odot}}

\newcommand{\zsun}{\ensuremath{Z_\odot}}

\newcommand{\sunits}{\mbox{\msun ~pc$^{-2}$}}
\newcommand{\pc}{\ensuremath{\mbox{pc}}}
\newcommand{\kms}{\mbox{km~s$^{-1}$}}
\newcommand{\xco}{\ensuremath{X_{\mathrm{CO}}}}

\newcommand{\aco}[1]{\ensuremath{\alpha_{\mathrm{CO}}}}
\newcommand{\co}[1]{\mbox{$^{#1}$CO}}

\newcommand{\xunits}{\mbox{cm$^{-2}$ (K km s$^{-1}$)$^{-1}$}}
\newcommand{\aunits}{\mbox{$M_\odot$ (K km s$^{-1}$ pc$^2$)$^{-1}$}}
\newcommand{\counits}{\mbox{K km s$^{-1}$}}

\newcommand{\arc}{\mbox{$^{\prime\prime}$}}

\newcommand{\mlum}{\ensuremath{M_{\rm lum}}}
\newcommand{\mvir}{\ensuremath{M_{\rm vir}}}
\newcommand{\mdyn}{\ensuremath{M_{\rm dyn}}}
\newcommand{\lco}{\ensuremath{L_{\rm CO}}}
\newcommand{\mstar}{\ensuremath{M_\star}}
\newcommand{\mhi}{\ensuremath{M_{\rm HI}}}
\newcommand{\hen}{\mbox{Henize 2-10}}

\newcommand{\mhtwo}{\ensuremath{M_{{\rm H}_2}}}

\shortauthors{Imara \& Faesi}

\begin{document}

\title{ALMA Observations of Giant Molecular Clouds in the Starburst Dwarf Galaxy Henize 2-10}

\author{Nia Imara}
\affiliation{Harvard-Smithsonian Center for Astrophysics, 60 Garden Street, Cambridge, MA 02138}
\author{Christopher M. Faesi}
\affiliation{Max Planck Institute for Astronomy, K\"{o}nigstuhl 17, 69117 Heidelberg, Germany}
\affiliation{Harvard-Smithsonian Center for Astrophysics, 60 Garden Street, Cambridge, MA 02138}
\email{nimara@cfa.harvard.edu}

\begin{abstract}
We present new $\co{12}(J=1-0)$ observations of Henize 2-10, a blue compact dwarf galaxy about 8.7 Mpc away, taken with the Atacama Large Millimeter Array.  These are the highest spatial and spectral resolution observations, to date, of the molecular gas in this starburst galaxy. We measure a molecular mass of $(1.2\pm 0.4)\times 10^8~\msun$~in Henize 2-10, and $75\%$ of the molecular gas mass is contained within the northern region of the galaxy near the previously identified young super star clusters, which has a projected size of about 300 pc.  We use the CPROPS algorithm to identify 119 resolved giant molecular clouds distributed throughout the galaxy, and the molecular gas contained within these clouds make up between 45 to 70\% of the total molecular mass.  The molecular clouds in Henize 2-10 have similar median sizes ($\sim 26$ pc), luminous masses ($\sim 4\times 10^5$ \msun), and surface densities ($\sim 180$ \sunits) to Milky Way clouds.  However,  Henize 2-10 clouds have velocity dispersions ($\sim 3 \kms$) about $50\%$ higher than those in the Milky Way.  We provide evidence that \hen~clouds tend to be in virial equilibrium, with the virial and luminous masses scaling according to $\mvir\propto\mlum^{1.2\pm0.1}$, similar to clouds in the Milky Way.  However, we measure a scaling relationship between luminous mass and size, $\mlum\propto R^{3.0\pm0.3}$, that is steeper than what is observed in Milky Way clouds.  Assuming \hen~molecular clouds are virialized, we infer values of the CO-to-\htwo~conversion factor ranging from 0.5 to 13 times the standard value in the Solar Neighborhood. Given star formation efficiencies as low as 5\%, the most massive molecular clouds in Henize 2-10 currently have enough mass to form the next generation of super-star clusters in the galaxy.
\end{abstract}


\section{Introduction}
Among the many reasons why dwarf galaxies are so fascinating is that the conditions for star formation in these galaxies are quite unlike local conditions in the Milky Way.  A star born in a dwarf galaxy---with its low stellar mass ($\lesssim 10^9\msun$) and often comparatively gas-rich interstellar environment \citep[e.g.,][]{Mateo_1998}---may have been reared much differently than a typical star born inside the spiral arm of a larger galaxy.  This is because galactic environment may play a substantial role in shaping the properties and evolution of giant molecular clouds \citep[GMCs; e.g.,][]{Hughes_2013, Colombo_2014, Utomo_2015, Sun_2018}, the massive, cold reservoirs of molecular gas that set the stage for the initial conditions of star formation. 

Within the family of dwarf galaxies, of particular interest are blue compact dwarfs such as Henize 2-10.  With their small sizes \citep[e.g.,][]{Thuan_1991}, high current star formation rates \citep{Fanelli_1988, Thuan_1999}, high surface brightness, and compact star-forming regions \citep[e.g.,][]{Kunth_1988, Papaderos_1996, Thuan_1999}, starbursting dwarfs represent extreme star-forming environments that may be emblematic of the Universe's earliest galaxies \citep{Sargent_1970}.  The compact dwarf irregular Henize 2-10 has captured the interest of astronomers for decades.  First discovered in an H$\alpha$ survey by \citet{Henize_1967}, Henize 2-10 was only later identified as an extragalactic system \citep{Kondratjeva_1972}.  In this paper, we present new observations of the molecular, star-forming environment of Henize 2-10, whose properties are summarized in Table \ref{table1}.  Our study is part of the broader effort to understand how galactic environment shapes star formation at its earliest stages.

Henize 2-10 is located in the constellation Pyxis at a distance of $8.7$ Mpc \citep{Tully_1988}.  The galaxy is undergoing a major starburst, with a star formation rate (SFR) of $1.9$ \msun~yr$^{-1}$ \citep{Reines_2011}. The starburst activity is concentrated in two regions, called ``A'' and ``B,'' following the nomenclature of \citet{Vacca_1992}.  The regions have a projected separation of $8\arc$ \citep[$\sim 340~\pc$;][]{Vacca_1992}.  These compact regions contain hundreds of supergiants and tens of super-star clusters \citep[SSCs;][]{Johnson_2000}, which are thought to be common in merging galaxies and which may be the precursors to globular clusters \citep[e.g.,][]{Harris_2003}.  The two regions have been the subject of numerous spectroscopic studies \citep[e.g.,][]{Vacca_1992, Johnson_2000}, and radio and infrared observations have revealed the presence of SSCs still enshrouded in interstellar material  \citep[][]{Kobulnicky_1999, Turner_2000}.  The high far infrared luminosity of the starburst led \citet{Johansson_1987} to suggest that Henize 2-10 is a merger of two dwarf galaxies.  On the basis of extended, tail-like features observed in both CO and \HI, a study by \citet{Kobulnicky_1995} lent further credence to this view.  More recently, \citet{Reines_2011} and \citet{Reines_2012} identified a low-luminosity active galactic nucleus (AGN) coincident with the dynamical center of the galaxy, making \hen~the first known dwarf galaxy to host a supermassive black hole.

\input{table_he210.tex}

All in all, with its intense star formation activity, high gas content \citep[e.g.,][]{Kobulnicky_1995}, and AGN candidate, \hen~may  be similar in some ways to the Universe's earliest star-forming environments.  High-resolution observations of this galaxy thus provide a unique window into the physical conditions thought to be prevalent at higher redshift.

A number of previous authors have observed Henize 2-10 in several millimeter lines, including carbon monoxide (CO), the most observationally accessible tracer of molecular gas, especially in external galaxies.  \citet{Baas_1994} and \citet{Kobulnicky_1995} detected CO(1-0) emission in \hen, with the latter study revealing an elongated "tail" of molecular gas in the south-east, suggestive of a galaxy merger.  Neither of these studies had sufficient angular resolution to identify GMC-scale structure in the interstellar medium.   Higher transitions of CO have also been detected \citep{Baas_1994, Kobulnicky_1995, Meier_2001, Bayet_2004, Santangelo_2009}.  Of these studies, only \citet{Santangelo_2009} had high enough angular resolution ($1\farcs9\times1\farcs3$) to begin revealing the complex spatial structure of the ISM in \hen~and estimating properties of the galaxy's largest GMCs.  However, these observations did not recover the elongated tail detected by \citet{Kobulnicky_1995}.

Tracers of high density molecular gas in Henize 2-10 were observed by \citet{Imanishi_2007}, \citet{Santangelo_2009}, \citet{Vanzi_2009}, and most recently by \citet{Johnson_2018}.  \citet{Johnson_2018} imaged and detected the HCN(1-0), HNC(1-0), HCO$^+$(1-0), and CCH(1-0) lines at a resolution of $1\farcs7\times1\farcs6$, finding no evidence that regions associated with SSCs have preferentially larger line widths.  From their investigation of the relationship between CO(2-1) luminosity and line width, \citet{Johnson_2018} suggested that Henize 2-10 molecular clouds either have radii of $\sim 8$-32 pc, or instead, they may be larger but---possibly because of a high-pressure environment---have enhanced line widths.

This work presents new \co{12}(1-0) observations of \hen~taken with the Atacama Large Millimeter Array (ALMA).  With the highest resolution and greatest sensitivity millimeter-wave data to date, we map the distribution of molecular gas in Henize 2-10 with unprecedented detail.  With our resolution of $\sim26$ pc, we also identify GMCs, which in the Milky Way and other galaxies tend to range in size from $\sim 20$ to $100~\pc$ \citep[e.g.,][]{Bolatto_2008, Heyer_2015}.  Our main goals are to (1) use this exquisite data set to identify molecular clouds; (2) characterize their properties; and (3) discover whether and why there are differences between stellar nurseries in Henize 2-10 and the Milky Way and other galaxies.  

In Section \ref{sec:observations} we provide an overview of our observations.  In Section \ref{sec:global} we describe the global properties of the molecular gas, including the dynamics.  In Section \ref{sec:results} we identify GMCs in \hen, characterize their properties, and compare these properties with those of molecular clouds in other galaxies.  We further discuss our results in Section \ref{sec:discussion} and summarize our conclusions in Section \ref{sec:summary}.

\input{table_observations.tex}

\section{Observations}\label{sec:observations}
We obtained ALMA Cycle 4 observations (project code 2015.1.01569.S; PI: Nia Imara) in the CO(1-0) line at 115.2712 GHz towards \hen~in July and August 2016. A single position ($08^{\rm h}36^{\rm m}15\fs20, -26\degr 24\arcmin34\farcs 0$ [J2000]) was observed.  The ALMA 12-m array was in configuration C36-4  for the two observing nights, with 36 antennas each night, arranged with baselines from 15 m to 1124 m, implying a minimum angular resolution of $0\farcs 48$ and a maximum recoverable scale of $6\farcs 7$ (at 115.27 GHz). The half-power beamwidth for the 12-m array was $50\farcs7$.  Table 2 summarizes the observing dates, conditions, and calibrators.

The ALMA Band 3 correlator was set up to have a velocity resolution of 244.141 kHz (0.637 \kms) and a bandwidth of 469 MHz, centered at 114.936 GHz to cover the \co{12}(1-0) line, adjusted for the galaxy's LSR velocity of $856$ \kms~\citep{Kerr_1986}.  

Our data were processed and imaged using the Common Astronomy Software Applications (CASA) package\footnote{\url{http://casa.nrao.edu}}.  The North American ALMA Science team used CASA version 4.5.3 to manually calibrate the data.  We summarize the data processing steps as follows.  First, there were basic flagging operations, including autocorrelation, shadowed antenna, and edge channel flagging.  Next, a system temperature ($T_{\rm sys}$) calibration table was generated and deviant $T_{\rm sys}$ measurements were flagged.  Then the antenna positions were calibrated, followed by atmospheric calibration using the Water Vapor Radiometer data.  Finally, the bandpass, flux, and gain calibrations were performed. 

We imaged our data in CASA using the \texttt{multiscale} CLEAN algorithm.  This method, which searches for emission at a range of spatial scales, has been shown to do well at recovering extended emission, reducing the depth of negative emission features, and eliminating low-level flux missed by standard CLEAN algorithms, which only clean point source scale emission \citep[e.g.,][]{Rich_2008}.  With the aim of measuring the CO-derived properties of molecular clouds, natural weighting was chosen in order maximize sensitivity.  We detected continuum emission with a total flux of about 25 mJy and subtracted this from the line emission in the visibility domain.  Our process for imaging the CO line took several steps: first, we first examined each of the 200 $1~\kms$ velocity channels around the systemic velocity of Henize 2-10 to determine where significant emission is arising from spatially.  Next, a mask was drawn around all the significant and coherent emission.  This mask was then used to deconvolve all image planes.  Finally, we applied a primary beam correction, though we note that since most of the emission is concentrated in the central portion of the image, the effect on our measured flux is minimal. The final data cube has voxels with dimensions of $0\farcs 1\times 0\farcs 1 \times 1~\kms$ and a spatial extent of $51\arcsec$. The synthesized beam was measured to be  $0\farcs67\times 0\farcs58$, corresponding to a physical resolution of $28\pc\times 24\pc$ at the distance to \hen. The rms sensitivity is $3.7$ mJy per $1~\kms$ channel.

\section{Global Properties of Molecular Gas}\label{sec:global}
\subsection{Dynamics}\label{sec:dynamics}

\begin{figure}
  \plotone{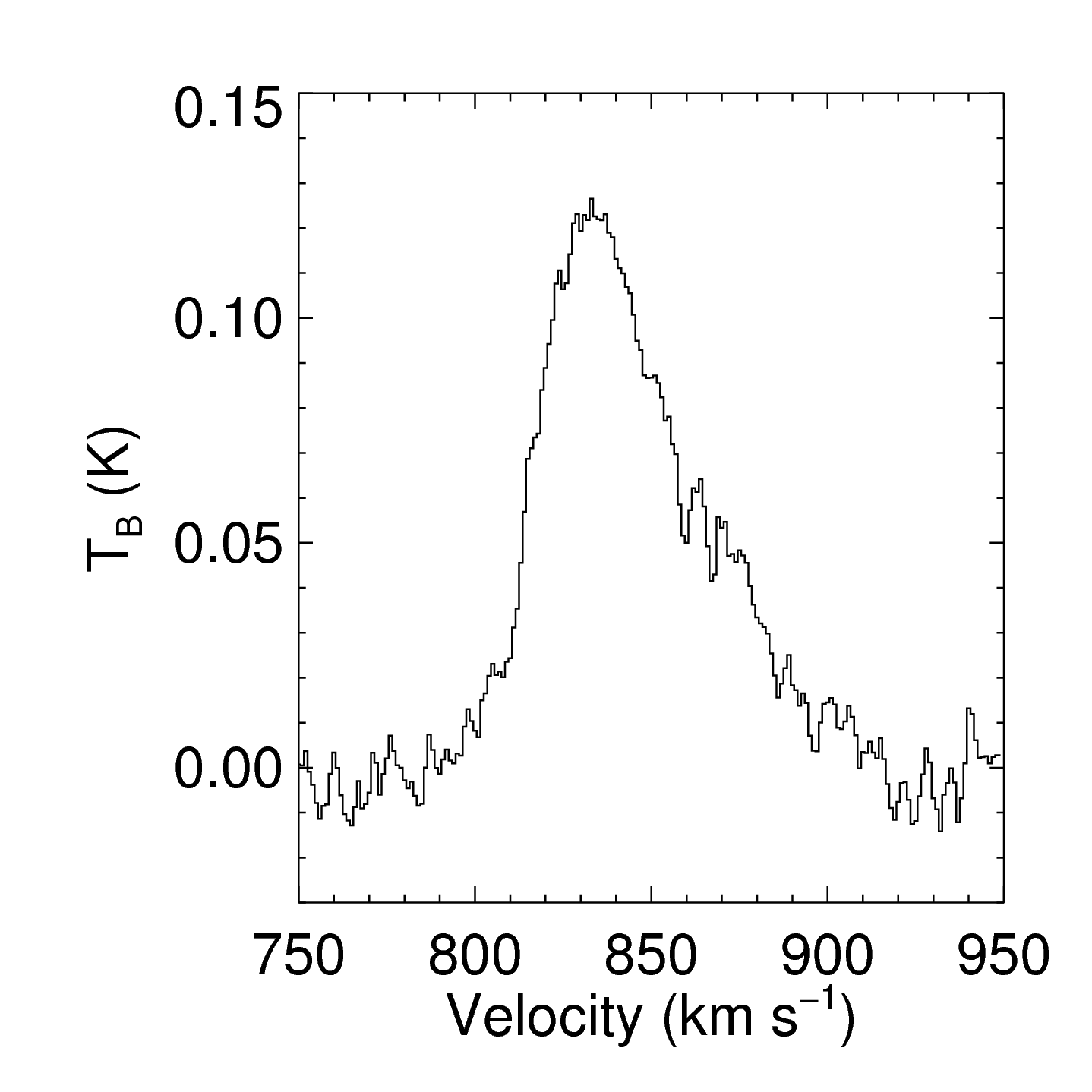}
    \caption{Composite \co{12} spectrum of Henize 2-10.}
    \label{fig:spectrum}
\end{figure}

\begin{figure}
    \centering
    \includegraphics[width=0.45\textwidth]{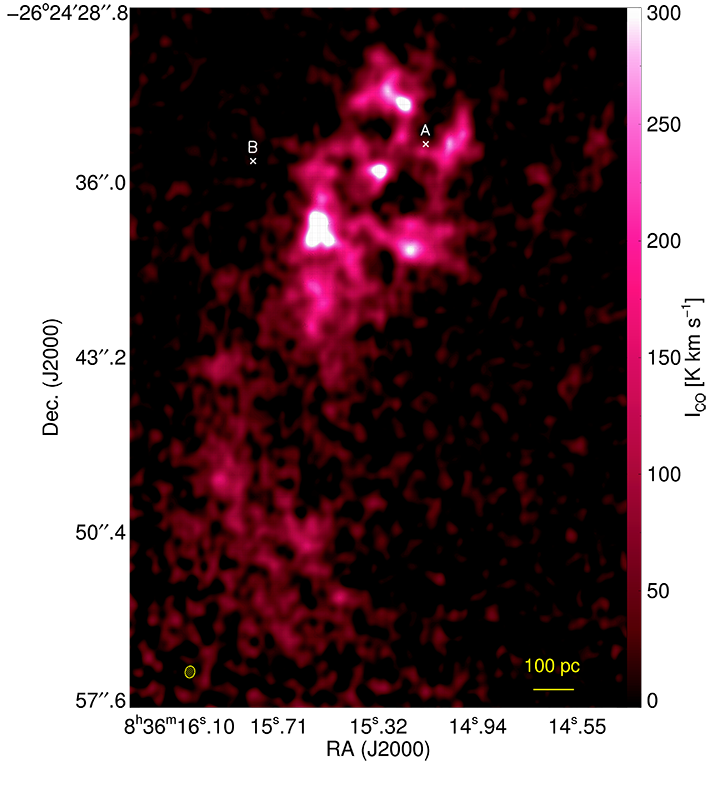}
    \caption{Total integrated intensity (zeroth moment) ALMA \co{12}(1-0) map of Henize 2-10.  The color bar gives the intensity in \counits.  The $0\farcs48\times 0\farcs58$ synthesized beam is indicated in the lower left.  The two $\times$ symbols represent the nominal centers of starburst regions A and B.}
    \label{fig:map}
\end{figure}

Figure \ref{fig:spectrum} shows the observed integrated \co{12}(1-0) spectrum of Henize 2-10 in units of brightness temperature, $T_B$, constructed simply by averaging $T_B$ across the image (displayed in Figure \ref{fig:map}) in each channel.   The profile is asymmetric, similar to those presented by \cite{Baas_1994}, \cite{Kobulnicky_1995}, and \cite{Santangelo_2009}.  

We fit a Gaussian to the spectrum, determining a systematic central velocity of the molecular gas of about 843 \kms, occurring at a peak temperature of $(2.37\pm0.03)$ K.  The fit gives a gas velocity dispersion of $24.60\pm 0.31$ \kms and a total integrated CO intensity of $I_{\rm CO}=146\pm 3$ \counits.

Figure \ref{fig:map} displays the \co{12} integrated intensity map, integrated over the velocity range 780 to 930 \kms.  The morphology of the molecular gas is complex, having a clumpy distribution, several ring-like structures, and high-intensity knots connected by diffuse regions.  We detect virtually no CO emission associated with starburst region B, while the nominal center of region A is situated near the edge of what appears to be a ring or arc of gas.

We note that the coordinates (J2000) of region A \citep[$8^{\rm h}36^{\rm m}15\fs14$, $-26\degr 24\arcmin34\farcs 0$;][]{Johnson_2000}, of the AGN \citep[$08^{\rm h}36^{\rm m}15\fs12$, $-26\degr 24\arcmin34\farcs 157$;][]{Reines_2012}, and of the nominal center of the galaxy \citep[$08^{\rm h}36^{\rm m}15\fs13$, $-26\degr 24\arcmin33\farcs 77$;][]{Evans_2010} are all nearly coincident.  These positions all coincide with the third of five ultra-compact radio knots identified by \citet{Kobulnicky_Johnson_1999}.  Moreover,  the brightest (H$\alpha$ equivalent widths greater than 100 \AA), youngest ($\lesssim 10$ Myr) SSCs identified by \citet{Johnson_2000} are in region A, which these authors interpret as the center of starburst activity in Henize 2-10.  Hereafter, we will refer to "region A" and "the center" of Henize 2-10 interchangeably.

\begin{figure*}
    \centering
    \includegraphics[width=6.25in]{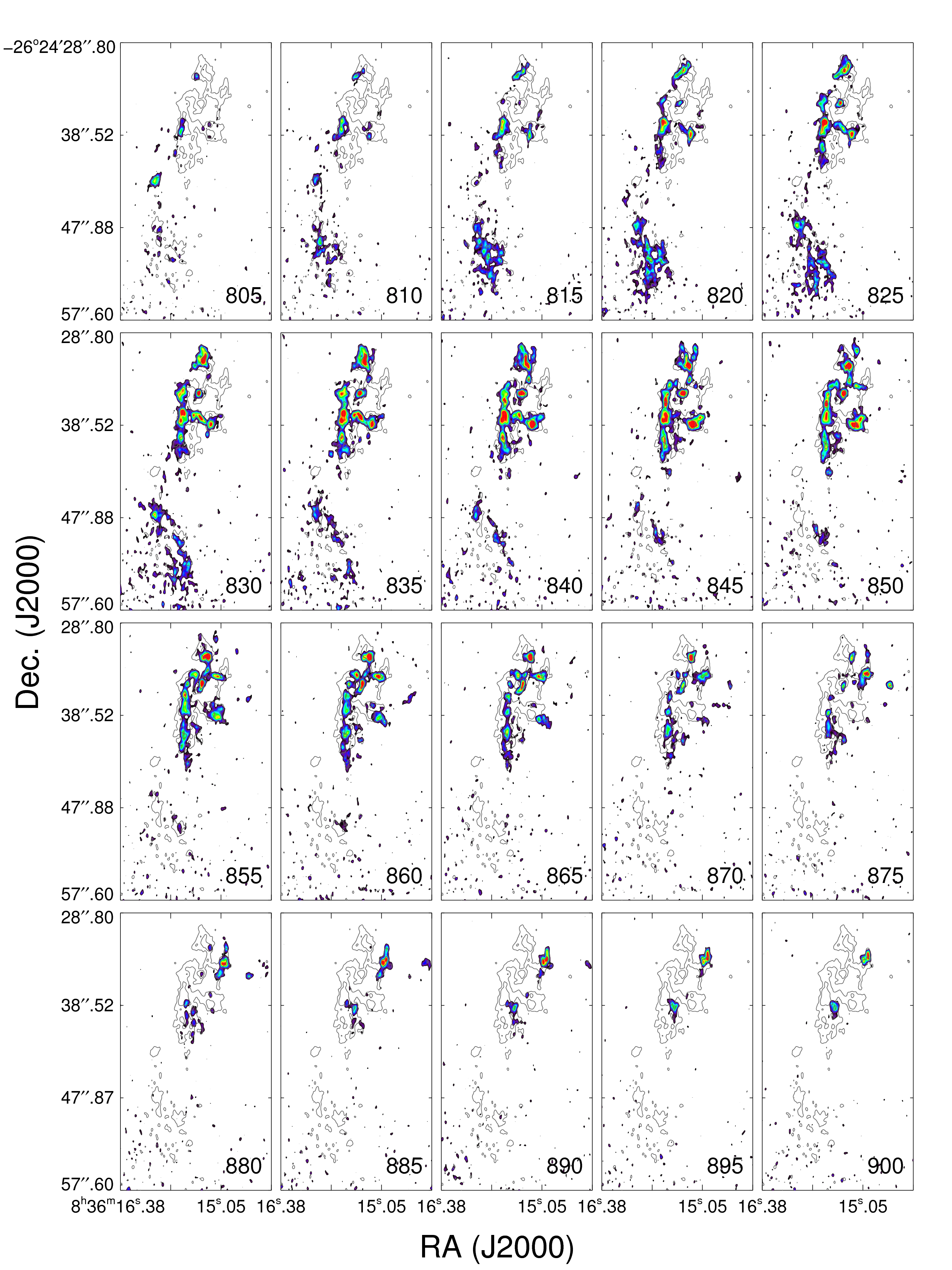}
    \caption{ALMA \co{12} channel maps with velocity widths of 5 \kms, shown in color.  The color scale goes from $0.50\sigma_{\rm rms}$ to $2.25\sigma_{\rm rms}$, where $\sigma_{\rm rms}=14.7$ \counits~is the rms noise level of the intensity map of the entire galaxy, integrated over the full velocity range.  For context, the contour map shows the structure of the entire galaxy integrated over all velocities.}
    \label{fig:channel}
\end{figure*}

In Figure \ref{fig:channel} we display twenty $5$ \kms-wide velocity channel maps, from 805 to 900 \kms.  The color scale in each channel goes from $0.5\sigma_{\rm rms}$ to $2.25\sigma_{\rm rms}$, where $\sigma_{\rm rms}=14.7$ \counits~is the rms noise level of the integrated intensity map of the entire galaxy (Figure \ref{fig:map}), integrated over the full range of velocities.  The bulk of the molecular gas emits strongly between about 820 to 875 \kms. The tidal feature appears at $\sim 805$-845 \kms~and reaches a peak intensity at about 820-825 \kms.  Arising prominently in the last row of channel maps are two compact regions of gas, located at $\rm{R.A.}=8^{\rm h}36^{\rm m}15\fs0$, $\rm{Dec.}=-26\degr24\arcmin33\farcs5$ and $\rm{R.A.}=8^{\rm h}36^{\rm m}15\fs4$, $\rm{Dec.}=-26\degr24\arcmin38\farcs6$.  These two knots are redshifted from the systematic velocity ($\sim 834~\kms$) at which the bulk of the molecular gas is moving.

\begin{figure}
    \centering
    \includegraphics[width=0.45\textwidth]{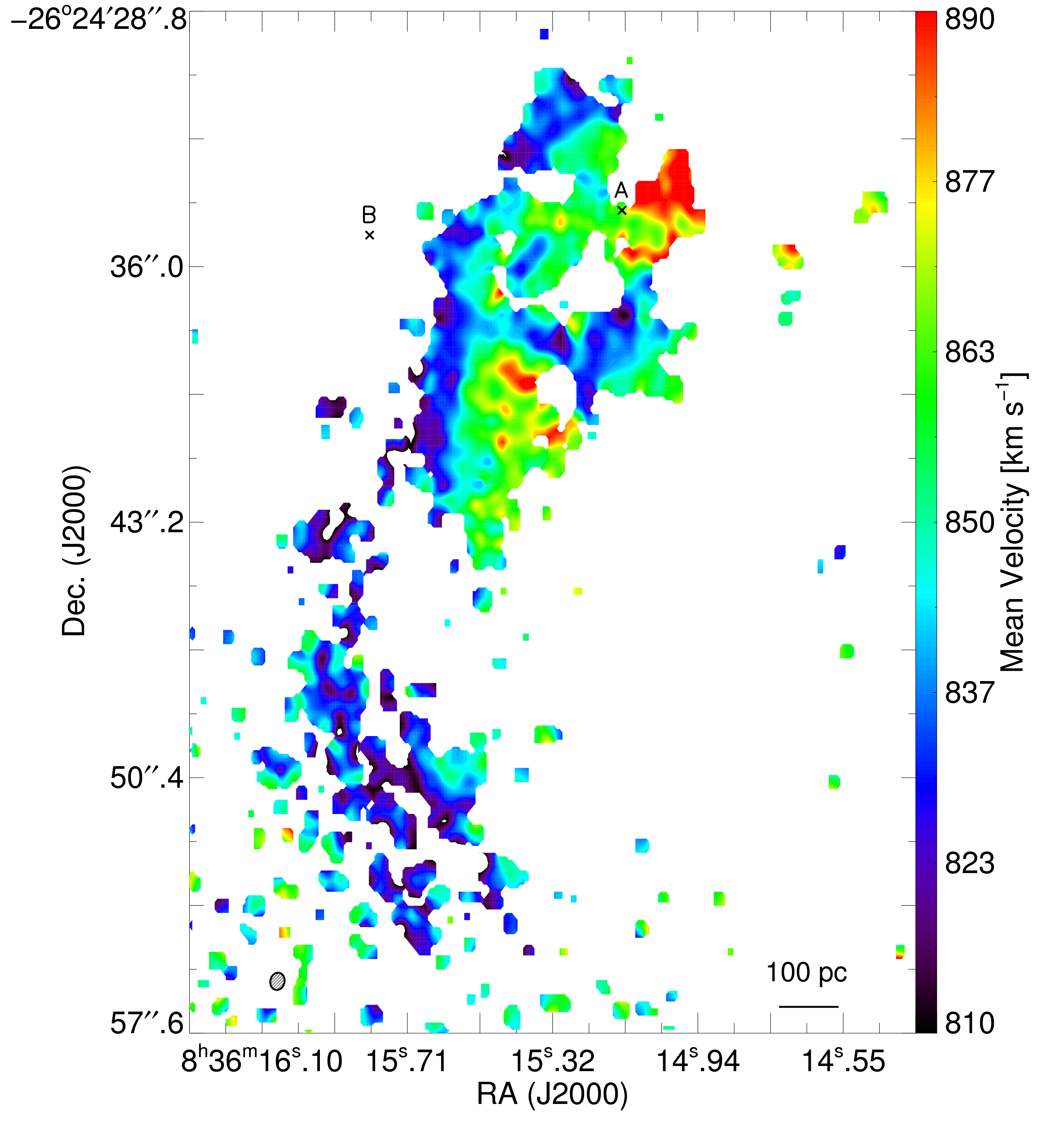}
    \caption{Intensity-weighted velocity (first moment) map of Henize 2-10.  The color bar represents LSR velocities in units of \kms.}
    \label{fig:m1}
\end{figure}

In Figure \ref{fig:m1} we present a map of the CO velocity field.  Individual channel maps used to create this intensity-weighted first moment map were clipped at $3\sigma$ (11 mJy/beam).  One immediately notices in Figure \ref{fig:m1} that the velocity field is not monotonic, as \citet{Kobulnicky_1995} also observed.  The same authors proposed that despite some variations, Henize 2-10 has an overall, relatively smooth velocity gradient that could be due to a large-scale molecular bar or disk. To assess the rotation curve in the inner regions of the galaxy near starburst A, \citet{Kobulnicky_1995} constructed a position-velocity diagram by taking a $4\arcsec$ wide slice passing through the center of starburst A at a position angle of $130\degr$, which they determine is the direction of the steepest gradient.  They then estimated a dynamical mass of $3.2\times 10^6$ \msun~within a 70 pc radius of the optical center.  Using our new ALMA data, we estimate the direction of the steepest gradient by fitting a plane to the intensity-weighted first-moment map of velocity centroids, following \citet{Imara_2011a}.  We perform the fit on the northern region of the galaxy, defined above.  In this manner, we estimate the angle of the steepest gradient to be $143\degr$.

\begin{figure}
    \centering
    \includegraphics[width=0.45\textwidth]{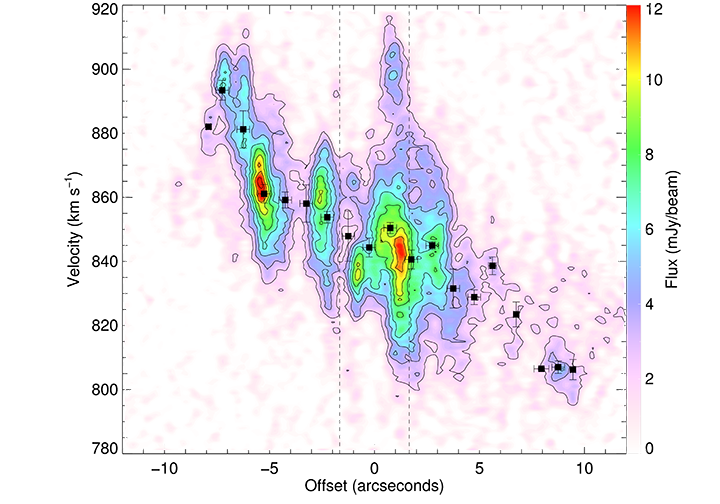}
    \caption{Position-velocity diagram of Henize 2-10 obtained from a $4\arcsec$-wide slice through the data cube along a position angle of $143\degr$, centered on starburst region A.  The contour levels go from 15\% to 85\% of the peak flux, in steps of 10\%.  Overplotted are the binned radial velocities as a function of radius (square symbols), in bins of $0\farcs5$.  The vertical lines correspond to a radial offset from the optical center of 70 pc.  The solid line is the best fit to the binned data within the inner 70 pc and has a slope of $-7.8$ \kms arcseconds$^{-1}$.}
    \label{fig:pv}
\end{figure}

In Figure \ref{fig:pv}, we display a position-velocity diagram constructed by taking a  $4\arcsec$ wide slice passing through the center of starburst A at a position angle of $143\degr$.  The velocity profile is roughly linear within $3\arcsec$ of starburst region A, indicating that the gas motions are schematically consistent with solid body rotation in this region.  To further emphasize the roughly linear trend, we also overplot the intensity-weighted radial velocities as a function of radial offset, binned every $0\farcs5$.  We perform a least-squares fit to the binned velocities within 70 pc ($1.7\arcsec$) of starburst region A, and we measure a velocity gradient of $7.8$ \kms $\arcsec^{-1}$, corresponding to a velocity difference of $\Delta V=13$ \kms~within the inner 70 pc.  This implies a dynamical mass, $\mdyn=(RV^2/G)/(\sin i)^2$---where $i$ is the inclination of the region with respect to the line of sight---of $2.7\times10^6$ \msun.  This is similar to the dynamical mass reported by \citet{Kobulnicky_1995}, who assumed a velocity difference of $14.1$ \kms, although their coarser resolution data could not resolve the inner 70 pc.  Since we assume edge-on rotation ($i=90\degr$), our estimate of the dynamical mass is a lower limit.  For instance, inclinations of $45\degr$ or $30\degr$ would yield dynamical masses of $5.5\times10^6$ and $1.1\times 10^7$ \msun, respectively.  We note, however, that if the gas dynamics are governed by phenomena other than rotation, such as supernova- or bar-driven outflows, or if the CO emission does not trace the gravitational potential of the stars, then the dynamical mass may be overestimated by our assumption of solid body rotation.

\citet{Johnson_2000} obtained UV spectra of starburst region A, taken with the \emph{Hubble Space Telescope} Goddard High Resolution Spectrograph.  Using population synthesis modeling, they inferred a total stellar mass of $1.6$-$2.6\times 10^6$ \msun.  \citet{Reines_2011} argued that the radio source in the region is likely due to an actively accreting supermassive black hole (SMBH) with a mass of $2\times 10^6$ \msun.  But \citet{Cresci_2017} recently disputed this conclusion, concluding from new MUSE data that an outflow from a starburst provides a better explain for the dynamics of the region.  Thus, if our interpretation of the velocity gradient is correct, then the star clusters and (potential) SMBH contribute most, if not all, of the total dynamical mass in the inner 70 pc around region A---if the rotating structure is edge-on.  But we know that the total molecular mass within 70 pc is $6.0\times 10^6$ \msun.  Thus, given the combined mass of the stars, black hole candidate, and gas in this region, we can constrain the inclination angle of the rotation to be no more than $31\degr$-$35\degr$.  Additional contributions to the total mass in this region---coming from, for instance, obscured stars, atomic gas, and dust \citep[e.g.,][]{Vacca_2002}---would result in lower values for the inferred inclination.

\subsection{Molecular Mass}\label{sec:molecular_mass}
We calculate the total luminous molecular mass $M_{\rm tot}$ by summing over all pixels in the integrated intensity map shown in Figure \ref{fig:map}, as follows:
\begin{equation}
M_{\rm tot} = \left(\sum I_{\rm pix} \delta x \delta y \right) D^2\times \xco \times 1.36 \times m_{{\rm H}_2},
\end{equation}
where $I_{\rm pix}$ is the CO integrated intensity of a pixel, the $\delta$ terms are the pixel sizes, $D$ is the distance,  $ m_{{\rm H}_2}$ is the mass of an \htwo~molecule, and the factor of 1.36 is included to account for helium.

Converting from CO intensity to \htwo~column density $N(\htwo)$ requires a CO-to-\htwo~conversion factor, 
\begin{equation}\label{eq:xfactor}
    \xco \equiv N(\htwo)/ I_{\rm CO}.
\end{equation}
Equivalently, $\alpha_{\rm CO}  = \mlum/\lco$.  In the Milky Way Galaxy, a typical conversion factor is $\xco = 2\times 10^{20}\xunits$ \citep[equivalently, $\alpha_{\rm CO}=4.35$ \aunits;][]{Dame_2001,Bolatto_2013}.  Assuming this value for Henize 2-10 yields a molecular gas mass of $(1.2\pm 0.4)\times 10^8~\msun$.  This value is in agreement with that obtained by \citet[$1.5\times 10^8\msun$ when scaled to 8.7 Mpc]{Kobulnicky_1995}, who used a larger CO-to-\htwo~conversion factor of $\xco = 3\times 10^{20}$ \xunits.  If we apply the larger conversion factor to our observations, we get $M_{\rm mol} = (1.8\pm 0.6)\times 10^8\msun$, still entirely consistent with \citet{Kobulnicky_1995}.

We also construct a masked version of the cube that should contain only robustly identified CO emission. To do this, we start with the non-primary beam corrected cube and seed the mask by identifying pairs of channels in a given pixel with signal $>5\sigma_{\rm pix}$, where $\sigma_{\rm pix}$ is the RMS noise for that pixel. We then expand the mask from these seeds to incorporate all emission connected to the initial maxima having intensities $>1.5\sigma_{\rm pix}$. Experiments with the exact values of these seeding and edge thresholds suggest that this particular combination best allows for inclusion of all real emission while excluding noise peaks and removing any remaining imaging artifacts. We then apply the primary beam correction to this masked cube, and create a masked integrated intensity image using the same procedure described above. Since this approach includes only robustly identified signal, this is the version of the integrated intensity map we use to calculate the average surface densities as discussed below.

The bulk of the molecular gas, about $75\%$ by mass, resides in the northern region of the galaxy, which we define as that west of $08^{\rm h}36^{\rm m}14^{\rm s}.96$ and north of $-26\degr24\arcmin 34.6\farcs2$. The mean integrated intensity of the north region is $50.4$~\counits, which corresponds to an average molecular gas mass surface density of $219$ \sunits~(assuming the Milky Way $\alpha_{\rm CO}$).

In the southern part of the galaxy, there is an extended stream of gas about $7\farcs6$ long, which corresponds to $320$ pc assuming a distance of 8.7 Mpc.  In the image of the \co{12} intensity map (Figure \ref{fig:map}), one sees that this extended stream is not associated with the starburst regions.  We define the stream as the remaining significant emission outside the northern region as defined above.  This stream is significantly more diffuse than the main part of the galaxy, having an average integrated intensity of $22.2$ \counits, corresponding to $\Sigma\approx 97\msun~\pc^{-2}$, assuming a Galactic CO-to-\htwo~conversion factor.  \citet{Kobulnicky_1995} first observed this feature and argued that it may be a tidal tail resulting from a merger of two dwarf galaxies.

Metallicity is expected to influence the CO-to-\htwo~conversion factor, with $\alpha_{\rm CO}$ decreasing as metallicity increases \citep[see][and references therein]{Bolatto_2013}.  A wide range of gas phase metallicities have been reported in the literature for Henize 2-10, reflecting the familiar problem of different metallicity calibrations leading to conflicting results \citep[e.g., ][]{Kewley_2008}.  \citet{Vacca_1992} used long-slit spectroscopic observations of starburst regions A and B, yielding sub-solar metal abundances of $12+\log(\rm{O/H})=8.06$ (0.23\zsun) and $8.61$ (0.83\zsun), respectively.\footnote{Metallicity is defined here as the abundance of oxygen with respect to hydrogen, $Z=\rm{O/H}$.  Following \citet{Asplund_2009}, we assume $(\rm{O/H})_\odot=4.90\times10^{-4};$ i.e., $12+\log(\rm{O/H})_\odot =8.69$.}  In contrast, \citet{Kobulnicky_1999} derived a super-solar metallicity of $12+\log(\rm{O/H})=8.93$ $(1.7\zsun)$ from the integrated emission-line spectrum of \hen.  \citet{Cresci_2017} conducted observations of Henize 2-10 with the Multi Unit Spectroscopic Explorer \citep[MUSE;][]{Bacon_2010} optical integral field spectrometer, and they used photoionization models to constrain multiple emission-line ratios simultaneously.  These authors derived a spatially resolved map showing a large metallicity gradient across the galaxy, with the central starburst region A having super-solar metallicity and the outer regions of the galaxy having metallicity as low as $12+\log(\rm{O/H})\approx8.3$

\citet{Esteban_2014} conducted observations of \hen~with a high-sensitivity echelle spectrograph on the \emph{Very Large Telescope}.  They took spectra in a $8\arc\times3\arc$ box in the brightest optical region of the galaxy, centered at ($08^{\rm h}36^{\rm m}15, -26\degr 24\arcmin33$ [J2000]).  In contrast to most of the other studies discussed above, \citet{Esteban_2014} detected the [\ion{O}{3}]4363 auroral line, which allows a determination of the electron temperature and thus the ionization parameter, and thus ultimately a better constraint on the oxygen abundance.  We take their resulting measurement of $12+\log(\rm{O/H})=8.55\pm0.02$ $(\sim 0.7\zsun)$ to be the most reliable in the literature and adopt this as the average metallicity for \hen, while acknowledging that the galaxy may have a metallicity gradient.  If Henize 2-10 does indeed have such a gradient, it is likely that the CO-to-\htwo~conversion factor varies across the galaxy as well.  In the case that the metallicity is sub-solar on average, implying a higher CO-to-\htwo~conversion factor the the typical Milky Way value we use here, it may be that we underestimate the molecular mass (though note that the conversion factor may also potentially be lower due to the starburst nature of the galaxy; see Section \ref{sec:virial} for more discussion).

\section{GMC Identification and Properties}\label{sec:results}
In this section, we describe how we identify individual molecular clouds in Henize 2-10, measure their properties, and compare them with clouds in other galaxies.  Our comparison sample was selected to include a wide range of galactic environments.  To provide as relevant a comparison as possible, we selected our sample from studies whose original observations have comparably high spatial resolution, high surface brightness sensitivity, and which used similar cloud identification algorithms to ours.  Note that throughout this paper we will use the terms "cloud" and "GMC" interchangeably.

\subsection{GMC Identification}\label{sec:identification}
To identify and characterize GMCs, we used the \texttt{CPROPS} algorithm of \citet{Rosolowsky_2006}.  This algorithm identifies significant emission within a three-dimensional data cube by searching for pairs of adjacent pixels with signal-to-noise greater than a user-defined multiple of $\sigma$, the rms noise level per channel.  It then includes all other pixels down to a lower threshold level and defines each such distinct set as an "island" of emission.  Finally, individual clouds are decomposed from within these islands by identifying significant local maxima and the emission uniquely associated with each such maximum, then pruning from this set of candidate clouds those that have projected area smaller than the synthesized beam, an insufficient contrast in peak brightness as compared with the nearest merge level, or whose properties are robust to changes in including or excluding nearby candidate clouds (see \citet{Rosolowsky_2006} for details).  

After performing several experiments using various parameters of the \texttt{CPROPS} algorithm, we ultimately used the default parameters to decompose our data cube into GMCs and estimate their properties.  In particular, we selected a minimum threshold signal-to-noise of $4\sigma$ to seed the initial mask, and expanded to include all pixels $>2\sigma$ to generate the initial set of ``islands.''  Clouds were decomposed from within these islands using a moving box of $15~\pc\times 15~\pc\times 2~\kms$.  To ensure all emission is assigned to a cloud, we used the modified \texttt{CLUMPFIND} \citep{Williams_1994} routine, which partitions the emission into the final set of local maxima after the pruning stage. The measured cloud properties were then corrected for the limited sensitivity of our observations by extrapolating to a contour of 0~K and corrected for the limited angular and velocity resolution of the data as described in \citet{Rosolowsky_2006}.  This final correction involves deconvolving the spatial beam from the size of the cloud by subtracting the RMS beam from the extrapolated spatial moments in quadrature.  Similarly, the channel width is deconvolved from the second moment of the cloud in the velocity dimension \citep{Rosolowsky_2006}. For a detailed look at the effects of varying \texttt{CPROPS} parameters, we refer the reader to \citet{Rosolowsky_2006} and \citet{Faesi_2018}.

\begin{figure}
    \epsscale{1.1}
    \plotone{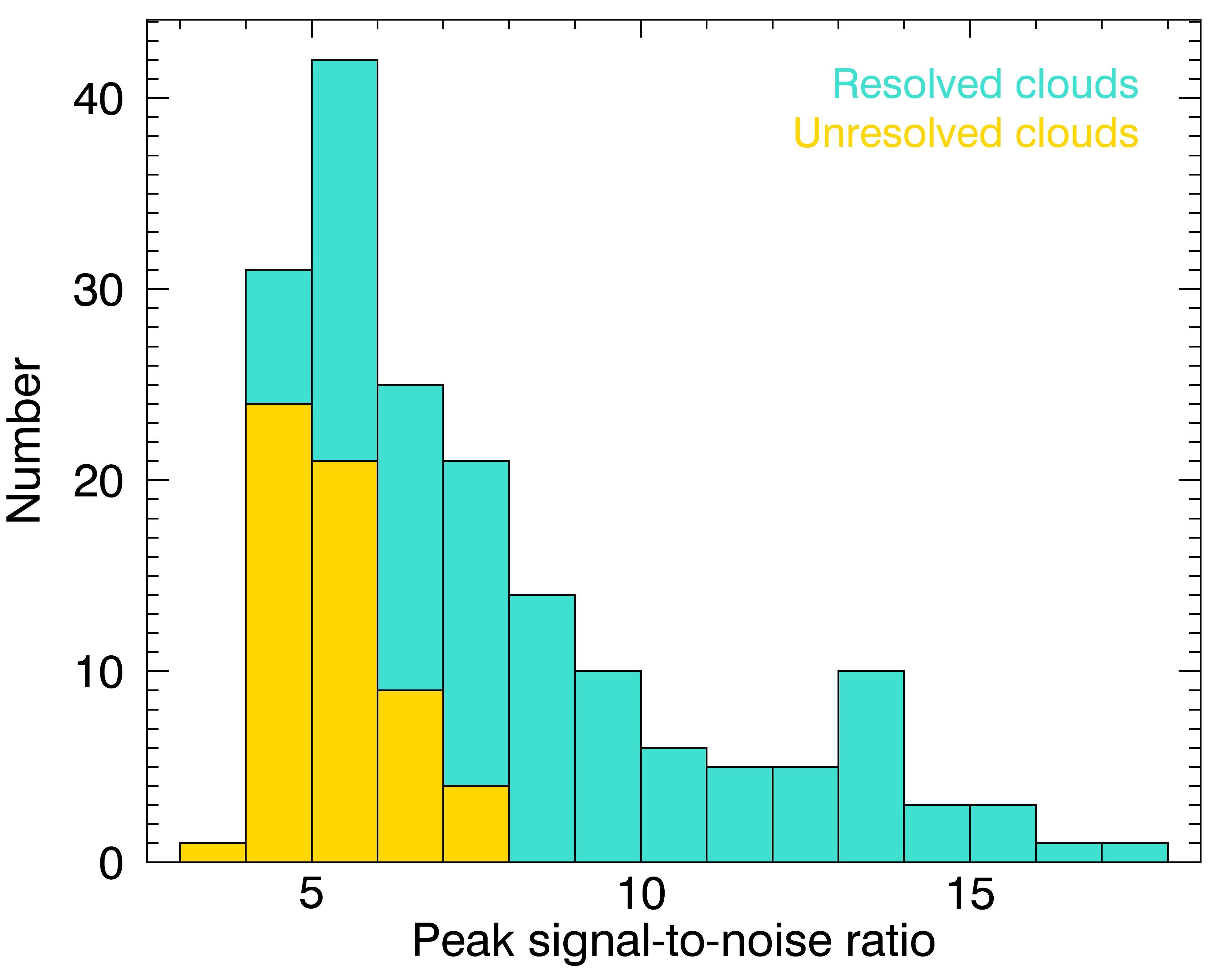}
    \caption{Histogram of the signal-to-noise ratio of the peak brightness temperatures of GMCs.  The turquoise and yellow bars represent resolved and unresolved clouds, respectively.}
    \label{fig:snr}
\end{figure}

\subsection{Distributions of Cloud Properties}\label{sec:properties}
Using \texttt{CPROPS} as described above we identify 178 clouds, of which 59 are unresolved, meaning that either their major or minor axis is smaller than the beam prior to deconvolution.  We take the 119 resolved clouds to be our final sample, the properties for which are listed in Table \ref{tab:cloudprops}.  A histogram of the signal-to-noise ratio of the peak brightness temperature of all clouds, both resolved and unresolved, is presented in Figure \ref{fig:snr}.

\begin{figure}
    \centering
    \includegraphics[width=0.45\textwidth]{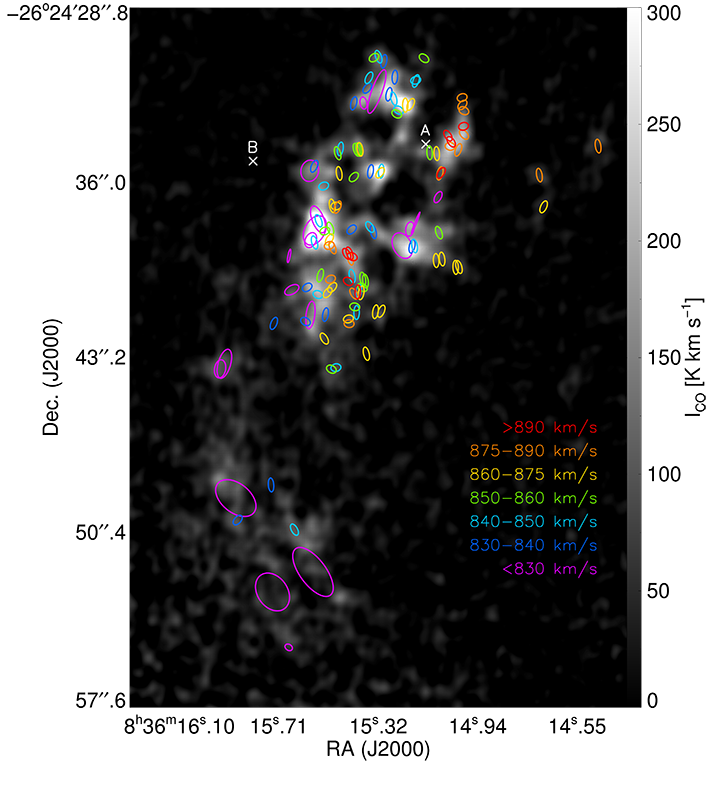}
    \caption{Total integrated intensity (zeroth moment) ALMA \co{12}(1-0) map of Henize 2-10 with GMC positions overlaid.  The ellipses represent the deconvolved sizes and position angles of the resolved GMCs.  The colors represent the velocity range of the GMCs' central velocities.}
    \label{fig:map2}
\end{figure}

\begin{figure}
    \epsscale{1.1}
    \plotone{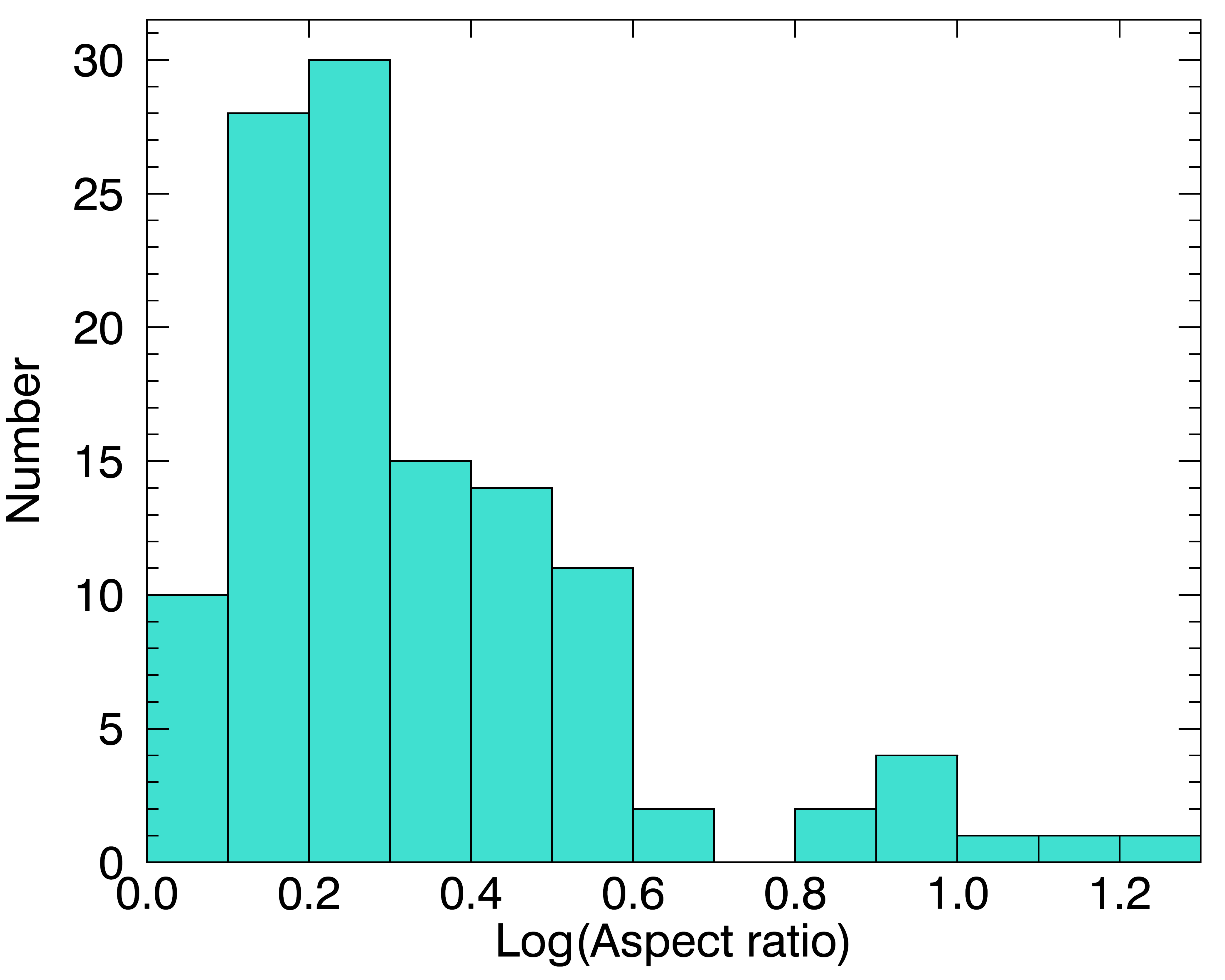}
    \caption{Histogram of the deconvolved aspect ratio of Henize 2-10 GMCs.}
    \label{fig:aspect}
\end{figure}

\begin{figure*}
    \centering
    \includegraphics[width=6.25in]{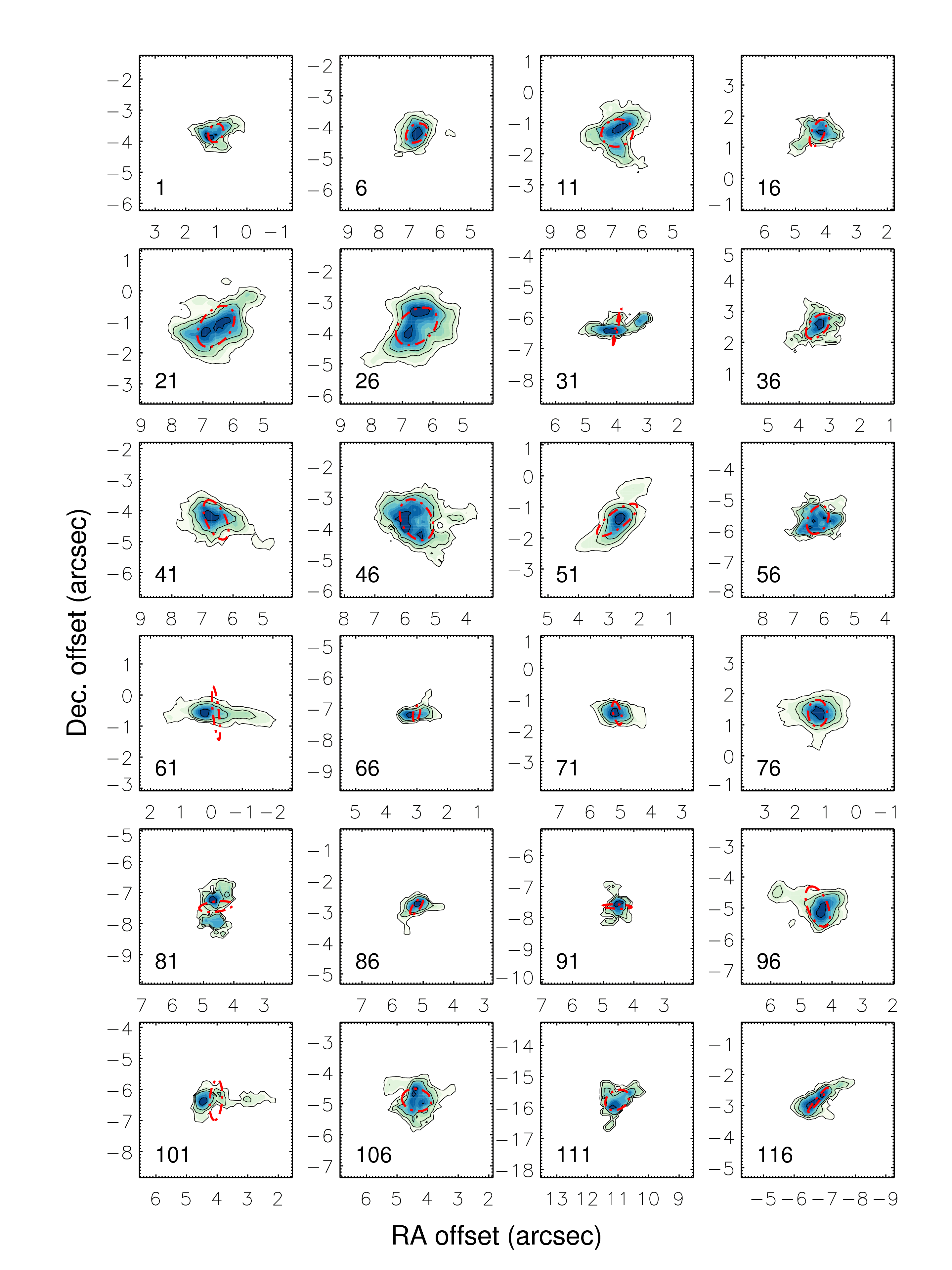}
    \caption{Integrated intensity maps of a sample of individual GMCs in Henize 2-10. Red ellipses show the deconvolved ellipses and position angles of clouds, and the numbers indicate the cloud identifications listed in Table \ref{tab:cloudprops}.  Each map is $5\arcsec\times5\arcsec$, and the axes display the angular separation between the center of a given GMC and the center of starburst region A.  The beam size is $0\farcs67\times 0\farcs58$.} \label{fig:multi_plot}
\end{figure*}
\begin{figure*}
    \centering
    \includegraphics[width=6.25in]{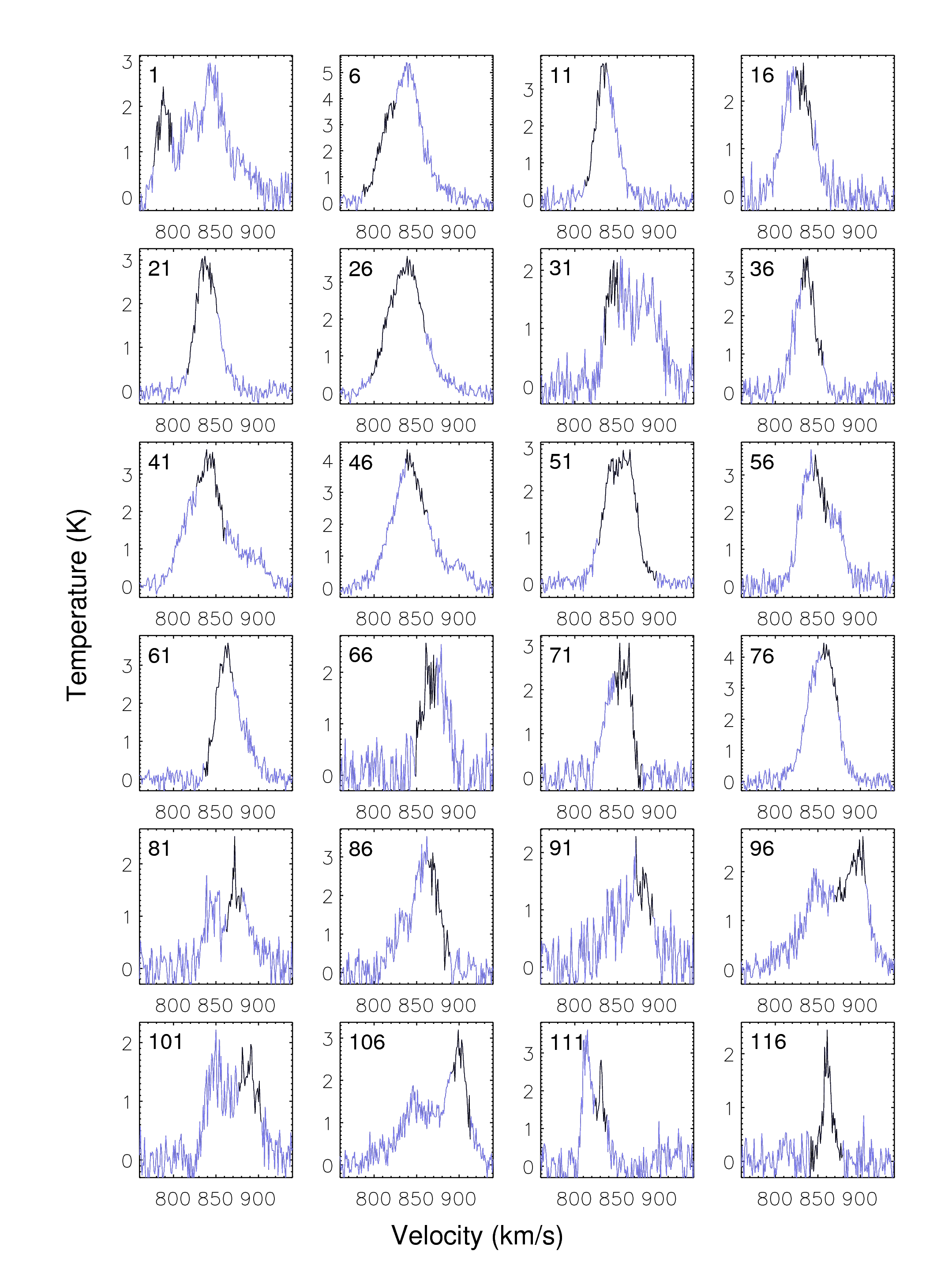}
    \caption{Average CO spectra toward the sample of individual GMCs displayed in Figure \ref{fig:multi_plot}.  The identification number of each GMC is printed in the upper left of the panel, and the velocities corresponding to that cloud are indicated in black on the spectrum. Features beyond that range corresponds to emission in a separate cloud.}   \label{fig:multi_spec}
\end{figure*}

Figure \ref{fig:map2} shows the CO integrated intensity map again, this time with ellipses indicating the FWHM sizes and orientations of the 119 resolved clouds.  Many of the clouds look slightly oblong, which becomes even more apparent when we calculate the GMC aspect ratios.  For each cloud, we define the aspect ratio as the deconvolved second moment of the emission along the major axis divided by the deconvolved second moment of the emission along the minor axis.  The median aspect ratios are 1.5 and 1.8 for resolved and unresolved clouds, respectively.  Figure \ref{fig:aspect} displays a histogram of the aspect ratios of all clouds.  Figure \ref{fig:multi_plot} shows the integrated intensity maps of a sample of individual GMCs in Henize 2-10.  In each panel of this figure, only emission from the isolated GMC is shown, and nearby clouds or clouds along the line of sight with different velocities are not displayed.  In other words, for each cloud we apply a 2D mask to each channel of the data cube and generate an intensity map constructed only from pixels assigned to that cloud.  Figure \ref{fig:multi_spec} displays the average spectra toward the clouds imaged in Figure \ref{fig:multi_plot}.  In each panel of this figure, emission from all clouds along the same line of sight toward an individual given GMC contribute to the average spectrum.

For clouds having projected radius $R$, one-dimensional velocity dispersion $\sigma_v$, and CO luminosity \lco, the virial mass, luminous mass,  and surface density are defined as follows:
\begin{equation}\label{eq:mvir}
    \mvir = 1040 R\sigma_v^2,
\end{equation}
\begin{equation}\label{eq:mlum}
    \mlum = \alpha_{\rm CO} L_{\rm CO},
\end{equation}
\begin{equation}\label{eq:surf}
    \Sigma = \frac{\mlum}{\pi R^2},
\end{equation}
where $R$ and $\sigma_v$ are in units of parsecs and km s$^{-1}$, and we assume the standard Milky Way value for \aco, the CO-to-\htwo~conversion factor. This calculation for \mvir~assumes spherical symmetry and a power law volume density ($\rho$) profile $\rho \propto R^{-1}$ \citep{Solomon_1987,Bolatto_2008}.

Uncertainties in the independent parameters $R$, $\sigma_v^2$, and \lco~are calculated using a bootstrapping method, which estimates errors by creating several trial samples from the initial cloud data \citep{Rosolowsky_2006}, with 500 iterations.  We compute uncertainties in the derived properties using standard propagation of error methods.

\begin{figure}
    \epsscale{1.1}
  \plotone{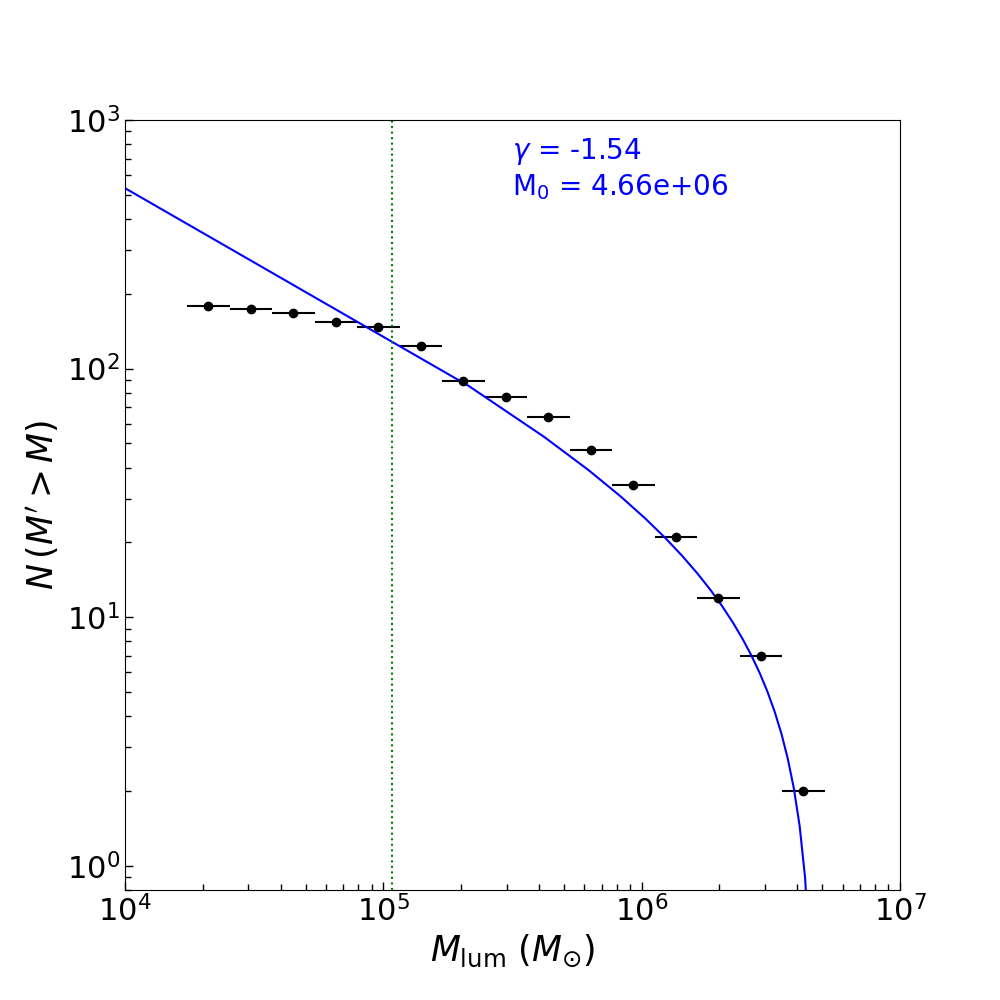} 
    \caption{Cumulative mass distribution of Henize 2-10 GMCs.  The solid line shows the best-fit truncated power law to the data above our completeness limit of $1.1\times 10^5\msun$.  We calculate a power law slope of $\gamma=-1.54\pm 0.10$, a truncation mass of $M_0=(4.7\pm0.8)\times 10^6\msun$, and $N_u=19.8\pm 9.0$ clouds.}
    \label{fig:cumulative}
\end{figure}
\subsection{The GMC mass spectrum in Henize 2-10}
The average luminous mass of our final sample of resolved GMCs is $7.0\times 10^5\msun$, and the median is $4.0\times 10^5~\msun.$  The mean and median of all clouds, both resolved and unresolved clouds, is $5.0\times 10^5\msun$ and $1.7\times 10^5\msun$  The mean and median masses of the \citet{Santangelo_2009} sample are $3.7\times 10^6$ and $3.1\times 10^6\msun$.  These values are likely overestimates, since their observations had coarser spatial and velocity resolutions, and because they had a lower sensitivity than ours, preventing the detection of smaller, lower mass clouds.

We next examine the distribution of GMCs by mass, or GMC mass spectrum, in \hen. Here we parameterize in terms of the cumulative mass distribution including an upper mass limit $M_0$ to the population \citep[e.g.,][]{Williams_1997, Rosolowsky_2005}:
\begin{equation}
    N(M^\prime > M) = N_u \left[\left( \frac{M}{M_0}  \right)^{\gamma + 1} - 1 \right],
\end{equation}
where $\gamma$ is the slope and $N_u$ is the number of clouds with mass greater than $M_u\equiv 2^{1/\gamma + 1}M_0$.

The cumulative mass distribution for \mlum~of GMCs in our final sample is presented in Figure \ref{fig:cumulative}.   We estimate a completeness limit of $1.1\times 10^5 \msun$.  We determine this number from the $3$-$\sigma$ observational sensitivity of our 0th-moment map ($\sigma=14.7$ \counits), which we convert to a luminosity, assuming a physical extent equal to the synthesized beam area ($\sim 559$ $\pc^2$) and a Galactic CO-to-\htwo~conversion factor of $4.35$ \aunits.  We fit the cumulative mass distribution to the data above the completeness limit using the \texttt{mspecfit.pro} IDL routine of \citet{Rosolowsky_2005}, which implements a maximum likelihood algorithm that incorporates uncertainties in the cloud mass to solve for $\gamma$, $N_0$, and $M_0$. For the \hen mass spectrum we derive a slope of $\gamma=-1.54\pm 0.10$, a truncation mass of $M_0 = (4.7\pm 0.8) \times 10^6$ \msun, and $N_u =  19.8\pm 9.0$ clouds.

Most observations of molecular-dominated regions in the Milky Way and nearby galaxies suggest that the bulk of clouds have low masses ($\gamma < 0$), and the majority of mass resides in high-mass clouds \citep[$\gamma > -2$; e.g.,][]{Williams_1997, Rosolowsky_2005, Kennicutt_2012}. In this respect, the Henize 2-10 GMC population is similar to those in the inner disks of the Milky Way \citep[e.g.,][]{Rice_2016}, M31 \citep{Rosolowsky_2007}, M33 \citep{Gratier_2012}, and M51 \citep{Colombo_2014}. Moreover, the high-mass cutoff and the number of clouds at the high-mass end of the distribution is similar to the inner Milky Way.  \citet{Williams_1997} compiled Milky Way GMC masses from \citet{Dame_1986}, \citet{Solomon_1987}, and \citet{Scoville_1987} and determined $M_0\approx 6\times 10^6$ \msun~and $N_u = 10$ \citep{Heyer_2015}.  The GMC mass spectrum slope in Henize 2-10 also demonstrates a remarkably similar slope to that measured in the Milky Way \citep[$\gamma=-1.59\pm 0.11$][]{Rice_2016}.  The mass spectra of less dense regions within galaxies, including the LMC \citep[$\gamma=-2.33$;][]{Wong_2011}, the outer disk of M33 \citep[$\gamma=-2.3$;][]{Gratier_2012}, and the interarm regions of M51 \citep[$\gamma=-2.5$;][]{Colombo_2014} have steeper slopes than that of Henize 2-10. This may reflect longer GMC growth timescales in these more quiescent regions as compared to inner galaxy disks or spiral arms \citep{Inutsuka_2015}.

To test the effects of the completeness limit on the fitting formalism, we also calculated the mass spectrum considering a completeness limit of $2\times10^5$ \msun (about twice our nominal completeness). We derive a slope of $\gamma = -1.70 \pm 0.16$, a truncation mass of $M_0 = (5.0\pm 0.9) \times 10^6$ \msun, and $N_u =  11.8\pm 6.8$ clouds, fully consistent with the results above. We thus consider our results robust to completeness effects above our sensitivity limit.

\begin{figure}
    \epsscale{1.1}
    \plotone{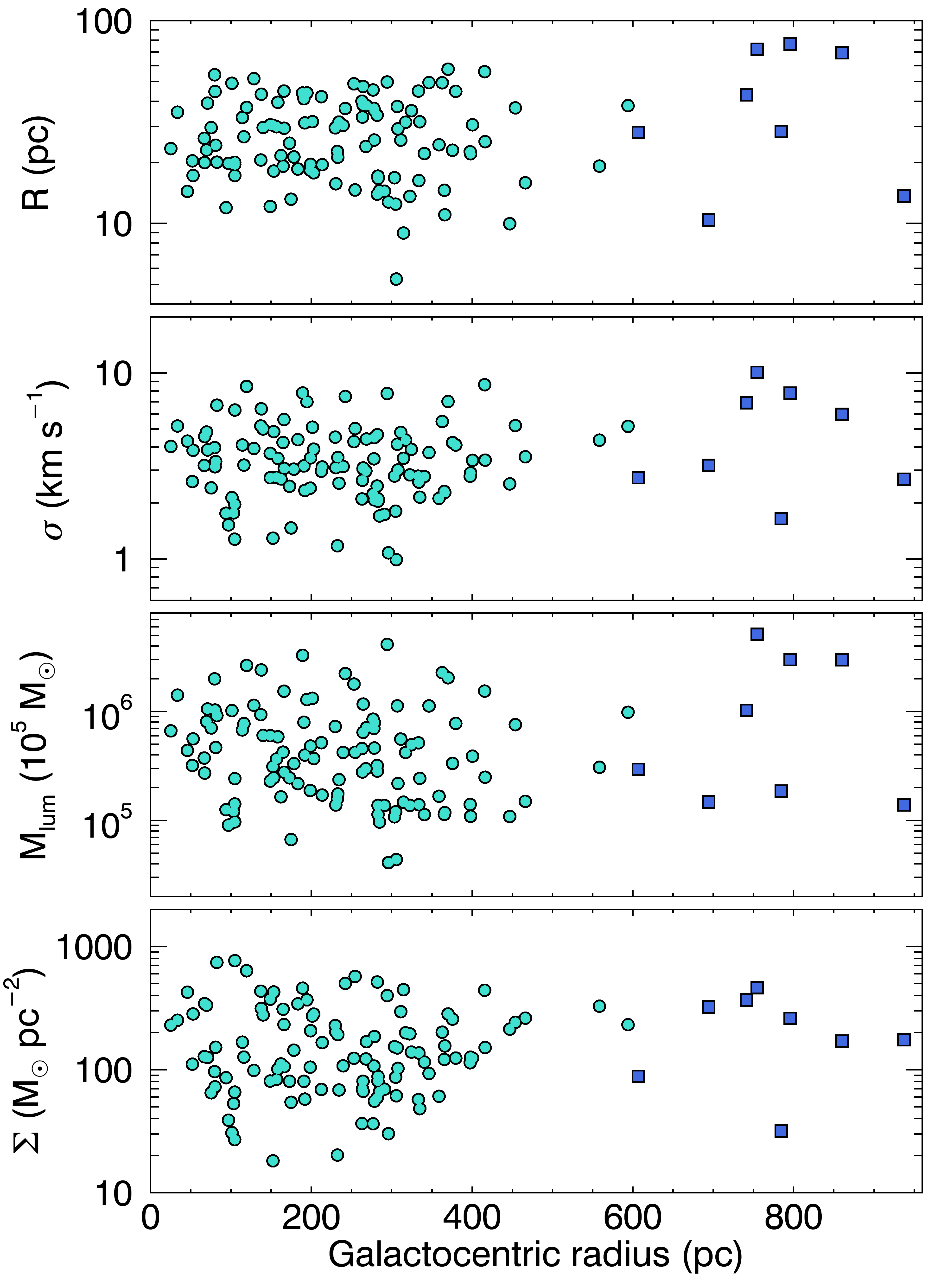}
    \caption{GMC properties as a function of distance from the center of Henize 2-10.  From top to bottom panel, size $R$, velocity dispersion $\sigma$, luminous mass \mlum, and mass surface density $\Sigma$.  GMCs associated with the extended feature in the south-east of the galaxy are indicated with dark blue square symbols.}
    \label{fig:distance}
\end{figure}

\subsection{GMC properties as a function of galactic environment}\label{sec:environment}
Figure \ref{fig:distance} shows GMC size, velocity dispersion, luminous mass, and surface density plotted as a function of the distance from the center of the galaxy, (corresponding to the position of starburst region A). These plots provide clues about how galactic environment may or may not influence GMC properties.  Though there are order of magnitude variations of these properties, there are no obvious systematic monotonic trends with distance.  We also experimented with binning the data, but this did not reveal any systematic trends either.  However, we do observe significant differences between the average properties of GMCs in the northern and southern regions of the galaxy.

As discussed in Section \ref{sec:dynamics}, the bulk of the molecular mass resides in the northern part of the galaxy, as do most of the GMCs, and there is a tidal feature in the south-east that may be the result of a merger.   One can see in Figure \ref{fig:distance} that most of the resolved molecular clouds (111 out of 119) are located within 600 pc of the galaxy center, while 8 clouds are associated with the south-east tidal ''tail."  On average, molecular clouds in the tail are larger ($43\pm7$ pc versus $28\pm5$ pc), more massive ($\mlum = (16.1\pm0.6)\times 10^5$ \msun~versus $3.6\pm0.7$ \kms).

What could explain these differences?  If the tail is indeed the result of a past merger \citep{Kobulnicky_1995}, then the higher velocity dispersions of clouds in this region could be due to tidal interactions. An increase in size and mass would balance the increase in velocity dispersion, such that these clouds are in virial equilibrium like their northern counterparts (see \S\ref{sec:larson}). It may also be that the spatial and kinematic structure of the molecular gas is more complex in the northern region, potentially due to the high energy and momentum input from the massive super star clusters in region A. This could naturally lead to a higher density of smaller, discrete structures that \texttt{CPROPS} identifies as separate clouds in the northern region as compared to the southern tail.  In addition, recall that \mlum~(and therefore, $\Sigma$), is calculated for all GMCs assuming a single, Galactic value for the CO-to-\htwo~conversion factor.  Thus, we cannot disentangle inherent variations in the luminous mass and surface density of clouds from potential variations in the conversion factor.

\begin{figure}[t]
    \epsscale{1.1}
    \plotone{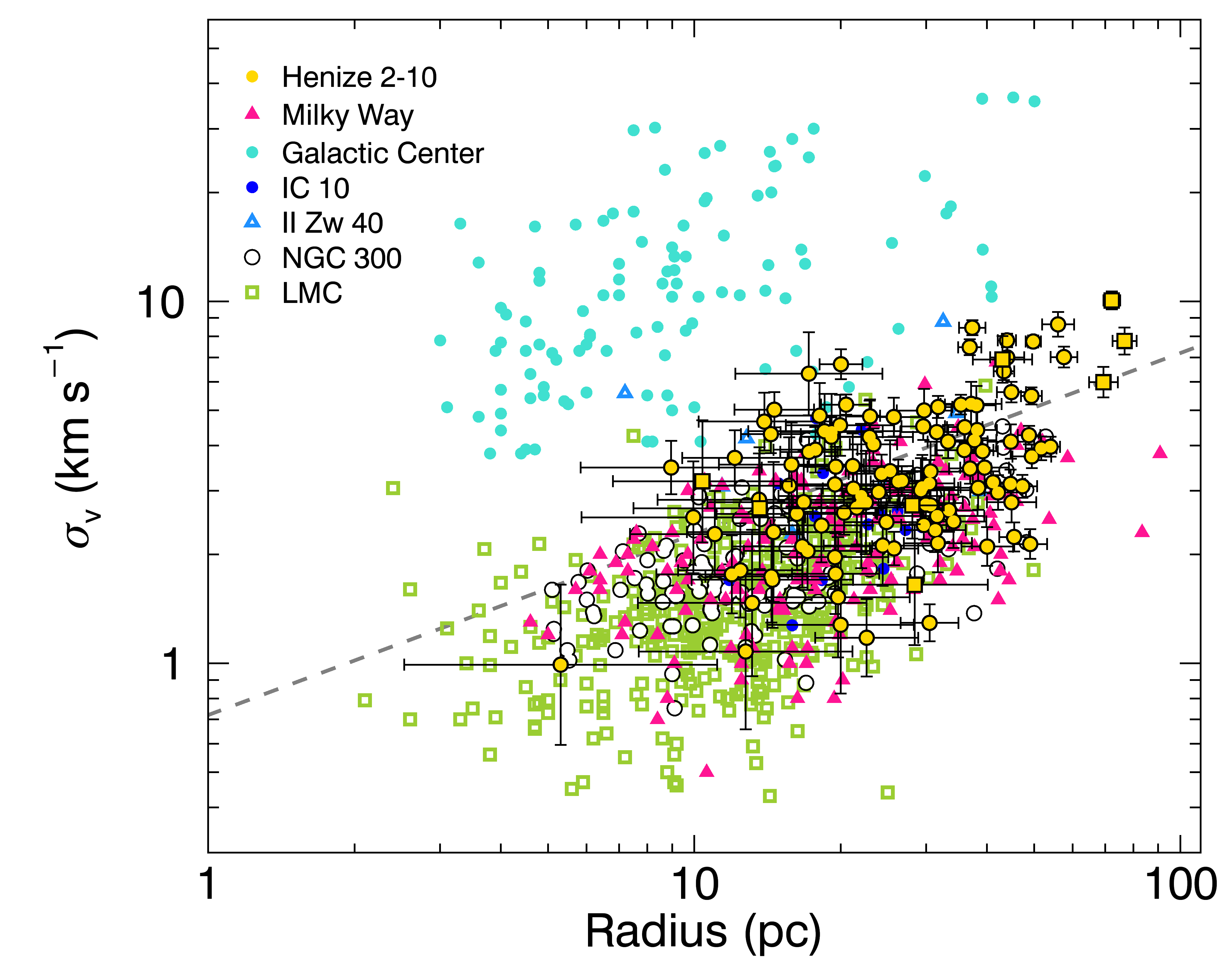}
    \caption{The size-linewidth relation for GMCs in Henize 2-10 and a comparison sample of clouds in other galaxies.  GMCs associated with the extended feature in the south-east of the galaxy are represented by square symbols.  The dashed line is the best fit to the sample of Milky Way GMCs as measured by \citet{Heyer_2001}.}
    \label{fig:size_linewidth}
\end{figure}

\begin{figure}[t]
   \epsscale{1.1}
  \plotone{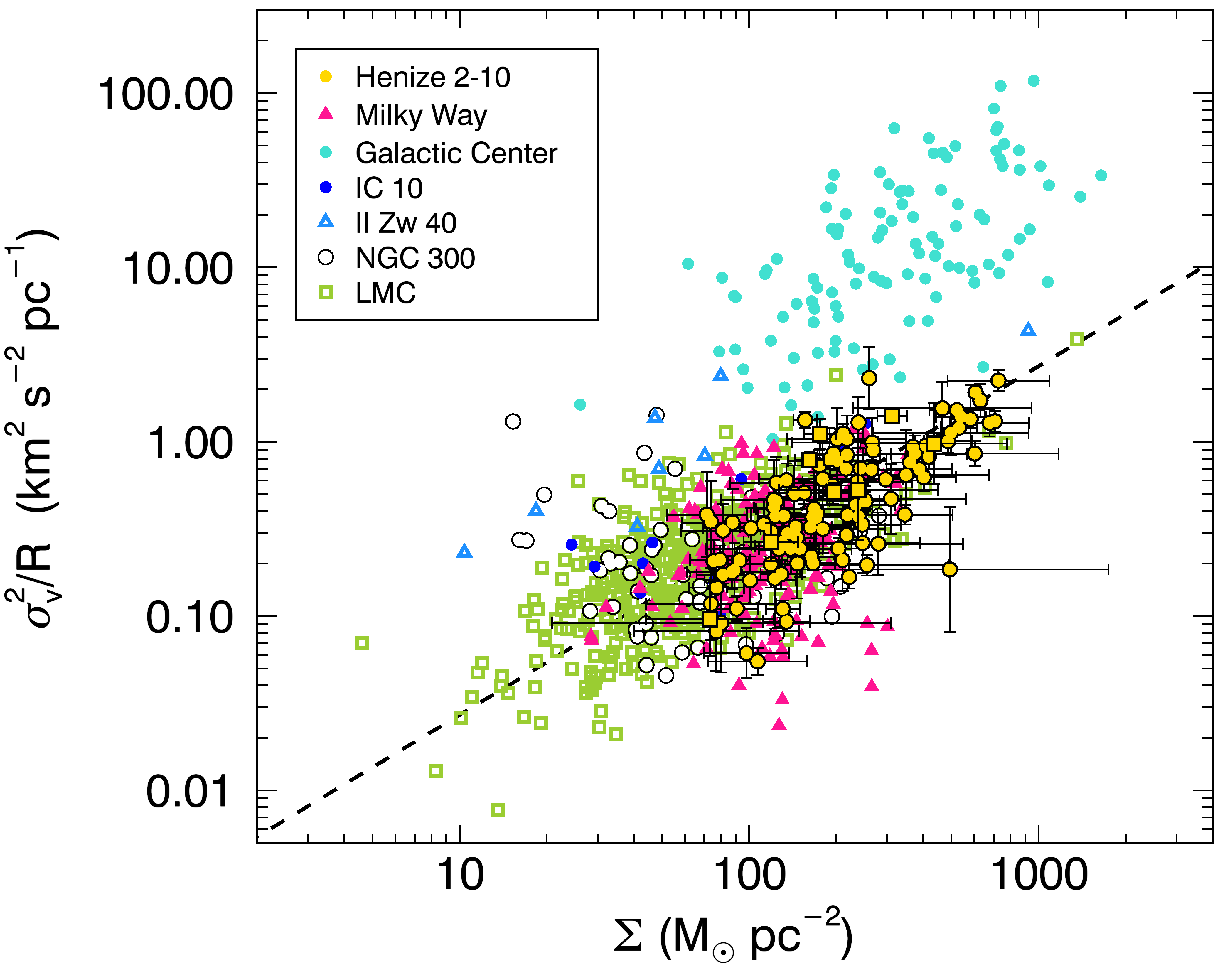}
    \caption{Size-linewidth coefficient as a function of mass surface density for GMCs in Henize 2-10 and a comparison sample of clouds in other galaxies.  $\Sigma$ is determined using \mlum, which is calculated assuming a Milky Way CO-to-\htwo~conversion factor. The dotted line represents the locus of virial equilibrium.}
    \label{fig:c_sigma}
\end{figure}
\subsection{Molecular Clouds in Other Galaxies}
To gain further insight into the physics of Henize 2-10 clouds, in the following section we compare their scaling relations to those of clouds in a sample of galaxies collected from the literature.  Here we describe the assembled literature cloud property measurements, which span a range of galactic environments.  \citet{Heyer_2009} made new measurements of the properties of Milky Way disk clouds from the \citet{Solomon_1987} catalog.  \citet{Oka_2001} studied Galactic Center clouds, observed  at a spatial resolution of 1.4 pc.  \citet{Wong_2011} measured cloud properties from CO-bright regions of the LMC at a resolution of 11 pc.  \citet{Faesi_2018} measured the properties of clouds in NGC 300, a spiral galaxy with a metallicity of roughly $0.6\zsun$ \citep{Deharveng_1988}.  \citet{Leroy_2006} investigated clouds in IC 10, a low-metallicity \citep[$\sim 0.3\zsun$][]{Magrini_2009}, at a resolution of 13.5-19.5 pc.  And \citet{Kepley_2016} measured the properties of clouds in II Zw 40, a starburst dwarf with a metallicity of $\sim 0.25\zsun$ \citep{Guseva_2000}, at a resolution of 24 pc.

Except for NGC 300 and II Zw 40, the molecular clouds in each of these galaxies were observed in the \co{12}(1-0) line.  \citet{Faesi_2018} observed CO(2-1) in NGC 300, and \citet{Kepley_2016} used CO(3-2) to identify clouds in II Zw 40. Both of these studies used standard assumptions for CO line ratios to derive cloud masses.

The assembled data sets come from studies that had different observational parameters (e.g., resolution and sensitivity). However, we note that all the extragalactic studies to which we compare utilized \texttt{CPROPS} to identify clouds. Since the Milky Way studies fully spatially resolve all objects, \texttt{CPROPS} was not necessary to identify GMCs in \cite{Heyer_2009} and \cite{Oka_2001}. It is beyond the scope of this paper to fully homogenize the methodology and data, though we refer the reader to \citet{Kepley_2016} and \citet{Faesi_2018} for studies that used \texttt{CPROPS} parameters most similar to our own. Finally, we also note that \texttt{CPROPS} is able to accurately recover the properties of model clouds similar to real objects in the Milky Way via simulated observations as if these clouds were at extragalacgtic distances \citep{Rosolowsky_2006}.

Our focus here is to broadly compare Henize 2-10 molecular cloud properties with literature data sets and to explore general similarities or dissimilarities between the cloud populations.  For Henize 2-10 and II Zw 40, we apply a Milky Way CO-to-\htwo~conversion factor of $\alpha_{\rm CO}=4.35$ \aunits~to estimate the luminous mass.    Following \citet{Leroy_2015}, we use a conversion factor twice this value for LMC clouds, according to the recommendation of \citet{Wong_2011}. Many studies have shown that Galactic Center clouds appear to have conversion factors 3--10 times smaller than disk clouds, likely due to dynamical effects and enhanced excitation \citep[see][and references therein]{Oka_2001, Bolatto_2013}.  For the Galactic Center clouds, we apply $\alpha_{\rm CO}=1$ \aunits, following \citet{Leroy_2015}.  For the \citet{Faesi_2018} cloud sample, we assume the luminous masses reported by these authors. 

\subsection{The Larson Scaling Relations}\label{sec:larson}
More than thirty years ago, \citet{Larson_1981} and \citet{Solomon_1987} used millimeter observations to demonstrate that Milky Way molecular clouds follow a set of well-behaved relationships between their basic properties, including size, velocity dispersion, luminosity, and virial mass.  Today, the so-called Larson's Laws are frequently used to compare extragalactic GMC populations to each other and to Galactic clouds.  The first relationship says that the linewidth of molecular clouds increases with size, according to $\sigma_v\sim R^{0.5}$ for Galactic clouds.  Typically called the size-linewidth relation, this is a statement that the internal turbulence of clouds increases as clouds get larger.  The second relationship states that CO luminosity (equivalently, the luminous mass) scales with the virial mass of molecular clouds.  Assuming that molecular clouds are virialized, this relationship is often used to determine a CO-to-\htwo~conversion factor.  The last relationship, stating that the mass scales with size, can be derived from the first two.  As a consequence of the first relationship and virial equilibrium, the usual physical interpretation is that all molecular clouds have roughly the same surface density.  However, GMC surface density variations have been observed between different galaxies or within distinct regions in individual galaxies, which may reflect local environmental conditions \citep[e.g.,][]{Colombo_2014, Sun_2018}. It is possible that GMCs in a given environment still have roughly constant surface density, with the precise value set by environmental facrtors \citep{Faesi_2018}.

\subsubsection{Size-linewidth relation}\label{sec:sizelw}
Figure \ref{fig:size_linewidth} displays the size-velocity dispersion relation for the resolved clouds in our sample plotted alongside clouds from other galaxies. Note that the velocity dispersion $\sigma_v$ differs from the FWHM linewidth by a constant factor of $\sqrt{(8~\ln{2})}$. Qualitatively, the Henize 2-10 clouds appear to occupy a similar region of this parameter space as those in the disks of the Milky Way and other galaxies (both dwarfs and spirals). In contrast, the center of the Milky Way appears to have much higher linewidth clouds for a fixed size than those in galaxy disks, including Henize 2-10. This may reflect the increased confining pressure in these environments; we will return to this point in \S\ref{sec:discussion}.

Figure~\ref{fig:size_linewidth} shows that Henize 2-10 GMCs in general seem more similar in size and linewidth to those in the Milky Way disk and other nearby galaxies than those in, for example, the Milky Way center. However, there is a large amount of scatter in our data beyond the error bars, suggesting that the size and linewidth may not be well-correlated in Henize 2-10. To explore this formally, we calculate the Pearson correlation coefficient $r_p$ between these quantities. We find that $r_p = 0.5$, with an associated p-value that rejects the null hypothesis of no correlation at a significance of $4\sigma$ (two-tailed). Size and linewidth are thus formally correlated in these GMCs in Henize 2-10, as is the case for GMCs in other galaxies, though the relatively low Pearson coefficient demonstrates that the correlation is not particularly strong.

We conduct a fit to the size and linewidth using nonlinear orthogonal distance regression, which incorporates uncertainties in both axes into the fit. The derived slope is $1.31\pm0.14$---much steeper than the value of $0.5$ that has been measured in the Milky Way and NGC 300 \citep[e.g.,][]{Heyer_2001,Rice_2016,Faesi_2018}.  Taken at face value, this could suggest that turbulent energy dissipation is more efficient for high sonic Mach numbers than low Mach numbers, a scenario suggested to qualitatively explain the steep size-linewidth relation seen in Milky Way Central Molecular Zone \citep{Kauffmann_2017}. Further testing would require detailed comparison to numerical simulations, which we defer to future studies.

Furthermore, we also note that we have not subtracted bulk motions from cloud linewidths, so linewidths for large clouds (which should show larger bulk velocity gradients) may be systematically overestimated, leading to an overestimation of the size-linewidth power law slope (i.e., a steeper derived slope than what is actually present in the data). Furthermore, the residual variance to the power law fit is 6.4, which, when considered alongside the relatively low Pearson coefficient, suggests that a power law is not actually a particularly good model for these data. We thus caution the over-interpretation of the exact value of this power law slope.

We note that most of the clouds in our sample have sizes, velocity dispersions, and luminous masses above the resolution and completeness limits. Based on their analysis of molecular clouds in NGC 300 at the limit of spatial resolution, \citet{Faesi_2018} discarded clouds that were less than half the size of the beam size of their observations.  Following \citet{Faesi_2018} in defining half the beam size as our spatial resolution completeness limit ($\sim 13$ pc), there are are only 9 resolved clouds in Henize 2-10 with deconvolved radii below this value.  There is only one resolved cloud having a deconvolved velocity dispersion less than the smoothed velocity resolution of $1$ \kms.  There are 9 clouds with luminous masses less than the completeness limit mass of $1.1\times 10^5$ \msun.  Thus, clouds lying below the completeness or resolution limits have a fairly minor effect on the Larson's relations.

To further assess how the difference in turbulence at a fixed size scale is related to the density structure of GMCs, Figure \ref{fig:c_sigma} compares the size-linewidth coefficient (which can be interpreted as the velocity structure function coefficient), $C=\sigma_v^2/R$, with the cloud surface density, $\Sigma$  \citep[see, e.g.][]{Heyer_2009}.

The dashed line in Figure \ref{fig:c_sigma} indicates the expectation for virialized clouds, $\Sigma\approx 331\sigma_v^2/R$. The Henize 2-10 GMCs scatter about this line, suggesting that they have linewidths that approximately balance their level of self-gravity. As compared to the LMC clouds and Milky Way disk clouds from the \citet{Heyer_2009} study, they are shifted slightly further upward along the $C$-$\Sigma$ relation. In particular, the Henize 2-10 GMC population median $\Sigma$ is larger by a factor of 60\% and factor of two than those of the Milky Way disk and LMC, respectively, while the Henize 2-10 GMC population median $C$ is larger by a factor $\sim2$ compared to the other two studies.

In the $C$-$\Sigma$ plane, deviations above and below the line reflect changes in the ratio of kinetic to gravitational potential energy (i.e., virial state), while changes in the direction along the line can be interpreted as differences in the ISM pressure in which the clouds are found \citep{Field_2011, Hughes_2013, Utomo_2015, Faesi_2018, Sun_2018}.  Our results thus suggest that the Henize 2-10 clouds reside in a slightly higher pressure environment than those in the Milky Way or LMC. It should be recalled that \mlum~is calculated for all GMCs---in the Henize 2-10 sample and the other galactic data sets, aside from NGC 300---assuming a single value within each galaxy for the CO-to-\htwo~conversion factor (see \S\ref{sec:environment}).  Thus, it is difficult to disentangle inherent variations in the surface density from potential variations in the conversion factor, which we do not account for (see \S\ref{sec:virial}). We do note that $\alpha_{\rm CO}$ varies among and within nearby galaxy disks by a factor of about 0.3 dex \citep{Sandstrom_2013}, and so this should be considered when comparing properties depending on CO luminosity (e.g., mass or surface density) between cloud populations. Measurements depending on velocity dispersion and size should be only minimally affected by systematic errors in the conversion factor, modulo effects of limited sensitivity to molecular gas mass at low metallicity.

\begin{figure}
 \epsscale{1.1}
  \plotone{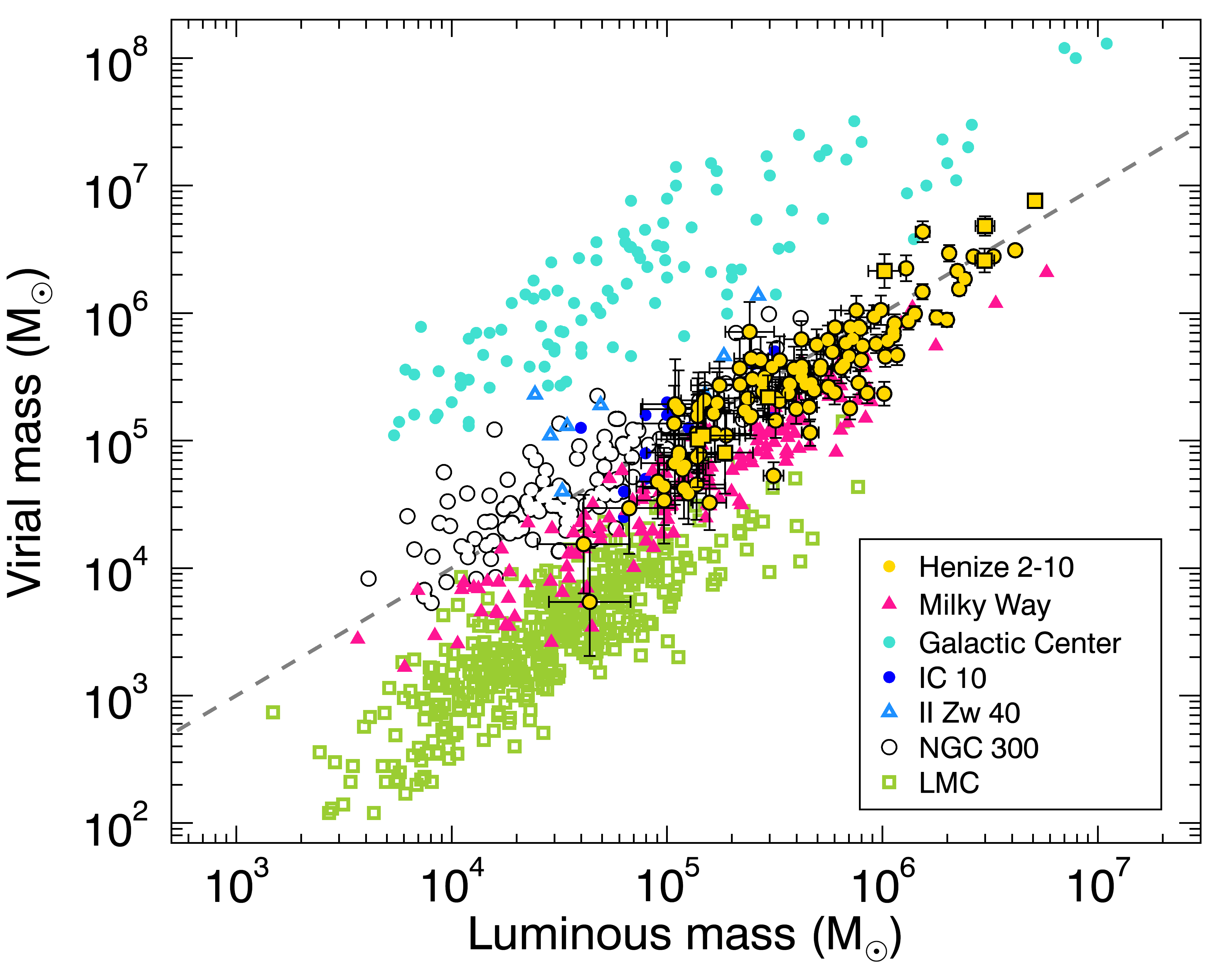}
    \caption{Virial mass versus luminous mass for Henize 2-10 GMCs and a comparison sample of clouds.  $\mlum$ is calculated for Henize 2-10 GMCs assuming a Milky Way CO-to-\htwo~conversion factor of $\alpha_{\rm CO}=4.3$ \msun~$[\counits \pc^2]^{-1}$.  The conversion factors for GMCs in other galaxies are summarized in the text. Henize 2-10 GMCs associated with the extended feature in the south-east of the galaxy are represented by square symbols.  The solid line indicates the one-to-one relation. Our data are well fit by a power law with slope $1.2\pm0.1$, shown by the dashed orange line.}
    \label{fig:virial}
\end{figure}

\begin{figure}
 \epsscale{1.1}
  \plotone{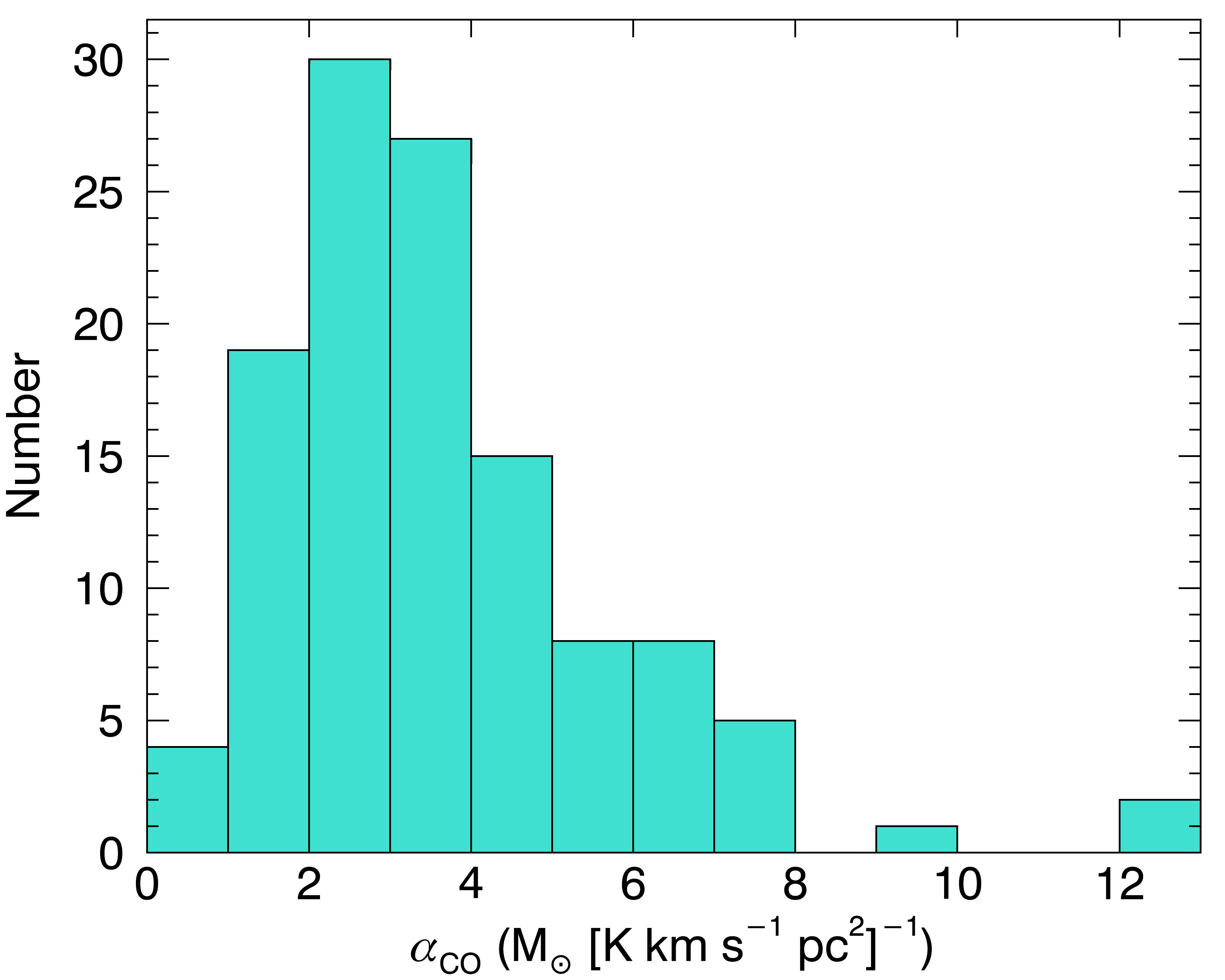}
    \caption{Distribution of derived CO-to-\htwo~conversion factors for individual GMCs in Henize 2-10, based on their virial masses and CO luminosities.  For comparison, the fiducial Milky Way value is $4.3$ \aunits~\citep{Bolatto_2013}.}
    \label{fig:xfactor}
\end{figure}

\subsubsection{Virial relation}\label{sec:virial}

Figure~\ref{fig:virial} shows the luminous mass versus virial mass estimates for the Henize 2-10 GMCs. The correlation is statistically tight, with a Pearson coefficient (in logarithmic space) of 0.90. The corresponding p-value is sufficient to reject the null hypothesis that virial and luminous mass are uncorrelated at the $10\sigma$ level. We perform a bilinear least-squares fit to the logarithm of the masses to derive a slope of $1.2\pm0.1$.  The clouds are on average near virial equilibrium (i.e., along the one-to-one locus). In this respect, Henize 2-10 GMCs are similar to clouds in the molecular ring of the Milky Way, which are also observed to be in virial equilibrium \citep{Larson_1981, Solomon_1987, Heyer_2009, Heyer_2015}.  The slightly superlinear slope implies that higher mass clouds are preferentially \textit{less} bound than lower mass ones, assuming CO is an equally faithful tracer of molecular gas at all masses. However, we note that (1) the majority of GMCs in our sample appear to be gravitationally bound (Figures~\ref{fig:c_sigma} and \ref{fig:virial}), and (2) our derived slope is within $2\sigma$ of unity, thus we do not further investigate this possibility.

Assuming the GMCs are in virial equilibrium, we can estimate the CO-to-\htwo~conversion factor as $\alpha_{\rm CO}=\mvir/\lco$.  This method of determining $\alpha_{\rm CO}$ assumes that the virial mass, \mvir, represents the actual molecular mass of a cloud.  The distribution of $\alpha_{\rm CO}$ derived under this assumption is displayed in Figure \ref{fig:xfactor}.  The implied conversion factor in Henize 2-10 clouds ranges from 0.5 to 13, and the median value is $3.7$ \aunits, slightly less than fiducial Galactic value of $4.3$ \aunits~\citep{Bolatto_2013}. 

If metallicity were the only influence on the CO-to-\htwo~conversion factor, and if indeed Henize 2-10 has on average a sub-solar metallicity, then one might expect the galaxy to have a higher conversion factor on average, compared to the Milky Way \citep[e.g.,][]{Bolatto_2013}.  On the other hand, the conversion factor in starbursts is expected to be driven down by high molecular gas temperature and velocity dispersion, which result from the merger activity and stellar potential in these systems \citep[e.g.,][]{Narayanan_2011, Bolatto_2013}.  Values as low as $0.8$ \aunits~have been adopted for starbursts \citep[e.g.,][]{Downes_1998}.  While \hen~molecular clouds do not display CO line widths significantly higher than Milky Way clouds, we cannot say anything about the gas temperature based on our observations. Since the metallicity and starburst effects on $\alpha_{\rm CO}$ described above operate in opposite directions, and we do not have sufficient direct constraints from our observations, we do not attempt to account for these effects in our calculations.

If the CO-to-\htwo~conversion factor depends on luminosity, as shown by previous authors for Milky Way clouds \citep[e.g.,][]{Solomon_1987, Bolatto_2013}, then our derived values for $\alpha_{\rm CO}$ may be a function of the luminosity range we sample.  However we do not observe a clear trend between $\alpha_{\rm CO}$ and luminosity for Henize 2-10 clouds, and so any changes in $\alpha_{\rm CO}$ due to luminosity may be minimized in this galaxy.  Ultimately, the values 3.7 and 4.3 are not all that different from each other; and so given the many uncertainties in determining  $\alpha_{\rm CO}$, given that $\alpha_{\rm CO}$ may vary across the galaxy, and for the sake of comparison with previous studies, we use a single, Milky Way value for the conversion factor of $\alpha_{\rm CO}=4.3$ \msun~$(\counits~\pc^2)^{-1}$ for this work.

\subsubsection{Mass-size relation}

\begin{figure}
    \epsscale{1.1}
  \plotone{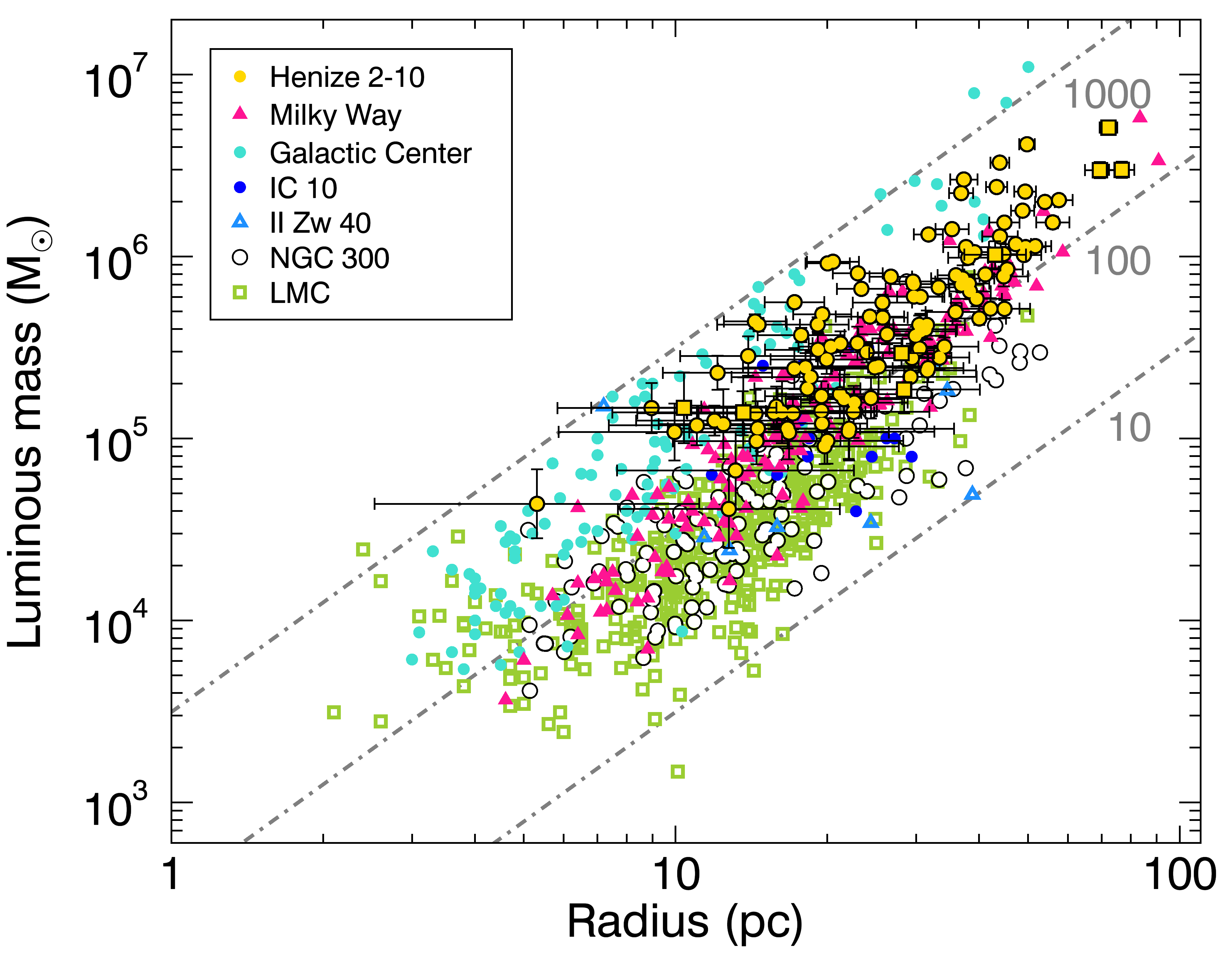}
    \caption{Luminous mass versus size for Henize 2-10 GMCs and a comparison sample of clouds. GMCs associated with the extended feature in the south-east of the galaxy are represented by square symbols.  Lines of constant surface density are shown for 10, 100, and 1000 \sunits. The best-fit line to the Henize 2-10 data, shown with the dashed orange line, has a slope of $3.0\pm0.3$.}
    \label{fig:mass_size}
\end{figure}

We present the relation between GMC luminous mass and size in Figure \ref{fig:mass_size} alongside a comparison sample from other galaxies. Henize 2-10 clouds occupy an intermediate region on the plot between Galactic Center clouds and NGC 300 clouds.  The median surface density is $180$ \sunits, similar to Galactic GMCs in the Molecular Ring \citep{Heyer_2015}, a point we return to in \S\ref{sec:discussion}. However, the precise scaling of the mass-size relation for Henize 2-10 clouds is quite different from what has been observed for Galactic clouds. 

Mass and size are closely correlated in the Henize 2-10 GMCs, with a Pearson coefficient of 0.83, and lack of correlation is ruled out at the $8\sigma$ level. A bilinear least-squares fit to the logarithms of the quantities yields a slope of $3.0\pm 0.3$.  This result stands in contrast to observations of Milky Way GMCs, for which \citet{Larson_1981} found an inverse correlation between volume density and size, implying that Galactic GMCs have roughly the same surface density, since $\Sigma\sim M/R^2\approx$ constant. This same scaling was also found in subsequent studies of Milky Way clouds \citep{Solomon_1987,Heyer_2009,Lombardi_2010} as well as in NGC 300 \citet{Faesi_2018}.  To test whether completeness effects alter our conclusions here, we calculated the best-fit slope for all Galactic clouds as well as for those only above the Henize 2-10 completeness limit of $1.1\times 10^5$ \msun.  For all Galactic clouds, the slope is 2.1., while for clouds above the completeness limit, the slope is 1.7.  Since for a given cloud size, Henize 2-10 has a larger number of more massive GMCs than does the Milky Way, this suggests that these more massive clouds drive up the slope in Henize 2-10.

In contrast to the Milky Way, our results suggest that instead of constant surface density, Henize 2-10 GMCs have constant volume density.  However, this is not a definitive conclusion, since the apparently steep slope assumes a constant CO-to-\htwo~conversion factor.  Nevertheless, while some Galactic data argue for constant surface density clouds, other studies suggest that environmental differences give rise to significant physical differences in surface density  \citep{Bolatto_2008, Heyer_2015, Utomo_2015}. As discussed in Section \ref{sec:sizelw} (see Figure~\ref{fig:c_sigma}), ISM pressure may be one such factor that sets the average surface density of GMC populations in different environments \citep{Faesi_2018}.

\section{Discussion}\label{sec:discussion}

\subsection{Comparison with the Milky Way}
A key goal of this study was to characterize the molecular gas and GMC properties in Henize 2-10 and to compare them with those in the Milky Way.  Our results suggest that despite these two galaxies having significant differences in their global properties---including mass, morphology, and specific star formation rate---their GMC populations have relatively similar mass distributions. The characteristic masses, sizes, and surface densities of the GMCs in these two galaxies are also similar. However, the Henize 2-10 clouds have higher linewidths than Milky Way clouds by a factor of 50\%, on average, suggesting increased turbulence (perhaps due to the starburst nature of \hen) as compared to the Milky Way disk.

\hen~has a stellar mass of $3.7\times 10^9$ \msun~\citep{Reines_2011}, an \HI~mass of $\mhi=3.1\times 10^{8}$ \msun~\citep{Sauvage_1997}, and a molecular mass of $1.2\times 10^8$ \msun (\S\ref{sec:molecular_mass}), each about an order of magnitude lower than the those in the Milky Way, which has stellar, \HI, and molecular masses of $5.2\times 10^{10}\msun$ \citep{Licquia_2015}, $8\times 10^9\msun$ \citep{Kalberla_2009}, and $(1.0\pm0.3)\times 10^9$ \msun~\citep{Heyer_2015}, respectively.  \citet{Reines_2011} determined the stellar mass for Henize 2-10 using measurements of the integrated K$_s$-band flux.  \citet{Nguyen_2014} analyzed the $g$- and $r$-band surface brightness profiles of Henize 2-10 and from these inferred a higher stellar mass of $(10\pm3)\times 10^9$ \msun, enclosed within 4.3 kpc.  In this paper, we adopt the \citet{Reines_2011} estimate for stellar mass, which is in close agreement with other estimates reported in the literature \citep[e.g.,][]{Kormendy_2013, Madden_2013}.

The SFR in \hen~is comparable to that of the Milky Way \citep[$1.65~\msun$ yr$^{-1}$;][]{Licquia_2015}.  Using far-infrared observations from \emph{Herschel}, \citet{Madden_2013} estimated $0.79~\msun$ yr$^{-1}$ for \hen, while \citet{Reines_2011} use H$\alpha$ and $24$ \micron~measurements to estimate an SFR of $1.9$ \msun~yr$^{-1}$.  With its lower stellar mass, however, \hen~has a specific star formation rate ($\rm{sSFR} = \rm{SFR}/\mstar$) from 6 to 16 times higher than the Galactic sSFR.  Note, if we assume a higher estimate of the stellar mass for \hen~\citep[$\sim10^{10}$ \msun;][]{Nguyen_2014}, its sSFR would be roughly 2--6 times higher than that of the Milky Way.

Henize 2-10 and the Milky Way have similar gas-to-stellar-mass fractions, $f_g\equiv (\mhi + \mhtwo)/\mstar$. For \hen~$f_g \approx 0.12$, while in the Milky Way $f_g\approx 0.17$.  However, \hen~has a greater proportion of its cold neutral gas in the form of molecular gas ($\sim 28\%$) than the Milky Way, which has about 11\% of its gas in the form of \htwo.  It follows that \hen~has a molecular gas depletion time-scale,  $\tau_{\rm dep}\equiv \mhtwo/\rm{SFR}$, of roughly 0.15 Gyr.  The molecular gas depletion time describes the amount of time it would take for a galaxy to use its entire supply of molecular gas at its current rate of star formation, assuming a closed system.   While the $0.15$ Gyr for \hen~is very similar to the molecular depletion time-scales of individual Milky Way GMCs in the solar vicinity \citep{Lada_2010, Lada_2012}, this is more than an order of magnitude faster than the $\sim 6$ Gyr depletion time in the Milky Way as a whole when all molecular gas is included.  This suggests that the current burst of star formation is probably short lived, as the galaxy will soon run out of star-forming fuel if it is not somehow replenished quickly.

\citet{Saintonge_2011} used \co{12}(1-0), optical, and UV observations of 222 galaxies having stellar masses in the range $10.0<\log\mstar/\msun<11.5$ and redshifts $0.025<z<0.05$ to investigate the molecular depletion time-scale.  They concluded that $\tau_{\rm dep}$ has a strong dependence on galaxy specific star formation rate and can be parameterized as $\log\tau_{\rm dep} = -0.44(\log\rm{sSFR}+10.40) + 8.98$, where $\tau_{\rm dep}$ and sSFR are in units of yr$^{-1}$.  For \hen, the sSFR determined from optical observations is $1.1\times10^9$ yr$^{-1}$, predicting a depletion time of $\tau_{\rm dep}\approx 0.23$, in agreement with our above results.  Thus, assuming the \citet{Saintonge_2011} relationship can be extrapolated to lower stellar masses, \hen~is most similar to the luminous infrared galaxies observed by these authors.

Not only does Henize 2-10 have a higher proportion of molecular gas than the Milky Way, that gas is at a much higher average density in comparison to the Galaxy.  \citet{Heyer_2015} estimated that the \htwo~mass within the solar circle---corresponding to Galactic radii between roughly 2 to 8.5 kpc---is $6.3$-$7.5\times 10^8$ \msun, depending on the Galactic model assumed. This corresponds to an average surface mass density within the solar circle of $3$-$3.5$ \sunits.  By comparison, the average surface density of molecular gas in \hen~is significantly higher, ranging from $97$ \sunits~in the southern tidal tail, to $219$ \sunits~in the northern part of the galaxy where most of the gas is concentrated (see  \S\ref{sec:molecular_mass}). Since the GMCs in these two galaxies have relatively similar properties, this large difference in average density could be due to two potential effects: either the molecular gas in Henize 2-10 has a much higher area (and presumably volume) filling factor than that of the Milky Way, or there is a large amount of relatively dense molecular gas not in GMCs in Henize 2-10. We explore these possibilities below.

\citet{Heyer_2015} reported the surface densities of GMC populations in the Galactic center, the Molecular Ring, and the outer Galaxy, with values of $1800$, $200$, and $30$ \sunits, respectively.  The surface densities, $\Sigma=\mlum/(\pi R^2)$, of Henize 2-10 clouds range from 71 to 727 \sunits, with mean and median values of 232 and 180 \sunits, respectively, similar to Galactic GMCs in the Molecular Ring. We note that \citet{Heyer_2009} examined the properties of Molecular Ring clouds in the \citet{Solomon_1987} sample using \co{13} data and derived a median surface density of $42$ \sunits.  In light of this particular study, Henize 2-10 GMCs seem to have much higher surface densities, on average, than Molecular Ring clouds.  However the surface densities determined by \citet{Heyer_2009} should be considered lower limits due to excitation effects and diminished \co{13} abundances in the low-column-density regime, and so we focus our comparison to the summary presented in \citet{Heyer_2015}.

The total luminous mass of the 119 cataloged GMCs in \hen~is roughly 67\% of the mass inferred from the total observed CO emission, compared to a value of 40\% in the Milky Way \citep{Heyer_2015}.  If we take into account the mass in all 178 clouds, including the unresolved clouds, this amounts to $\sim 71\%$ of the total molecular mass in the galaxy. Since the GMC surface densities are similar between these two galaxies, and Henize 2-10 has a higher fraction of molecular gas in GMCs than the Milky Way, the higher observed average molecular gas surface density in Henize 2-10 is primarily a result of a higher area filling factor as compared to the Milky Way. This may also explain why the molecular gas depletion time in Henize 2-10 is as short as that in Milky Way GMCs: the background molecular gas density is similar to the average GMC density, and so star formation can proceed across the galaxy in a manner similar to how it does in individual Milky Way GMCs.

There are some caveats to consider when evaluating the robustness of the fraction of molecular mass in GMCs.  If the \texttt{CPROPS} cloud identification algorithm missed some emission that should have been assigned to individual GMCs, this would lead to underestimates of their mass.  However, we remind the reader that we used the modified \texttt{CLUMPFIND} parameter (see \S\ref{sec:identification}), which ensures that all detected emission is assigned to a detected cloud.  Moreover, in measuring cloud properties, \texttt{CPROPS} extrapolates the sizes of clouds to 0 K to account for the effects of finite sensitivity.  On the other hand, if we have not accounted for all diffuse, extended emission in the calculation of the total molecular mass, this would result in an underestimate of the fraction of molecular gas locked up in GMCs, though we expect this effect to be minor since our total derived molecular mass is in agreement with the single dish observed value of \cite{Kobulnicky_1995}.

Extrapolating the GMC contours to 0 K raises another potential problem, because the integrated intensity of the entire galaxy has not been corrected for limited sensitivity.  Extrapolation introduces more emission to clouds than they would otherwise have and, therefore, leads to higher mass clouds. In Table \ref{tab:cloudmass} we provide estimates of the total luminous mass in GMCs, both extrapolated and not extrapolated, for the total sample as well as resolved clouds only.  If we compare the total unextrapolated mass in resolved clouds to the total mass of \hen, this lowers the estimated fraction of molecular mass within \hen~clouds to about $45\%$. We note that this reflects a rather strict lower limit on this fraction since unextrapolated cloud masses are systematically underestimated in simulated observations at our observed signal-to-noise \citep{Rosolowsky_2006}, and since we do not see evidence for significant extended CO emission missed in our observations.

\input{table_cloudmass.tex}


\begin{figure}
    \epsscale{1.1}
  \plotone{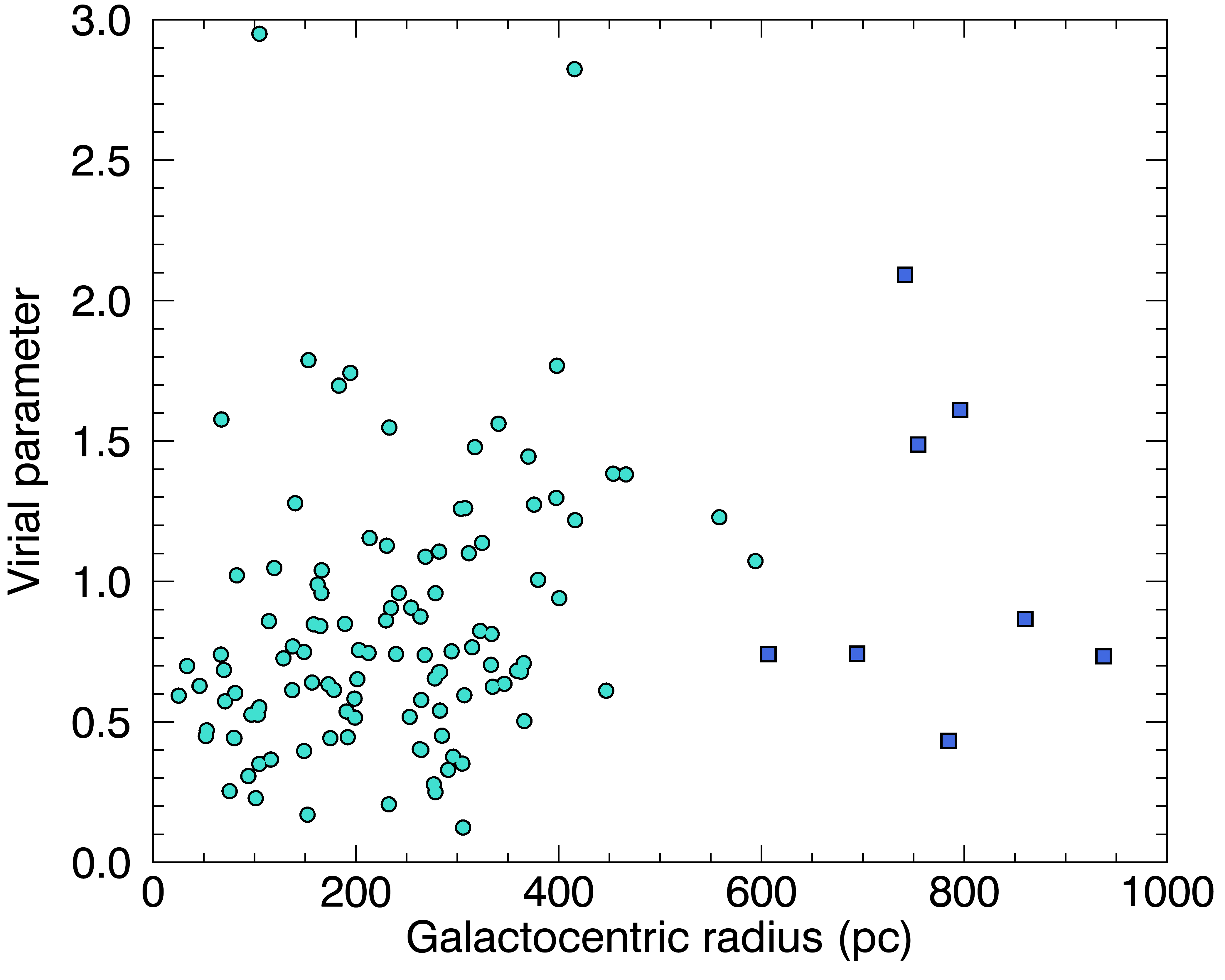}
    \caption{Virial parameter, $\mvir/\mlum$, as a function of projected distance from the center of Henize 2-10, (corresponding to the position of region A).  Within the main body of the galaxy, GMCs closer to the center tend to have smaller virial parameters than those further away.  The dark blue square symbols represent GMCs in the tidal feature.}
    \label{fig:virial_distance}
\end{figure}

\subsection{The Relationship Between GMCs and SSCs}\label{sec:gmcs_sscs}
In Section \ref{sec:environment} we investigated the possibility that galactic environment influences GMC properties.    We found that proximity to starburst region A does not have a significant impact on size, velocity dispersion, mass, and surface density.  Neither do we measure significant differences in the virial parameters of clouds in close proximity to starburst region A.  In \S\ref{sec:virial} we found a tight, nearly linear correlation among the GMCs between \mvir~and \mlum.  The virial parameter, $\alpha_{\rm vir}=\mvir/\mlum$, is frequently used to parameterize the relationship between gravitational and kinetic energy in physical systems, with values near unity indicating virial equilibrium \citep{Bertoldi_1992}.  In a cloud with $\alpha_{\rm vir}\lesssim1$, gravity is more important than kinetic energy, and the cloud may be supported by magnetic fields, if it is not collapsing on a free-fall timescale.  In a cloud with $\alpha_{\rm vir}\gtrsim1$, kinetic energy dominates over gravity, and the cloud must be confined by external pressure if it is not expanding and dispersing.  

For Henize 2-10 clouds, the mean and median values of  $\alpha_{\rm vir}$ are 0.85 and 0.74, respectively, with $71\%$ of the clouds having $\alpha_{\rm vir}<1$.  Given the likelihood of a metallicity gradient in Henize 2-10 \citep[][see \S\ref{sec:molecular_mass}]{Cresci_2017}, which would suggest variations in the CO-to-\htwo~conversion factor across the galaxy, the corresponding calculations for \mlum~and $\alpha_{\rm vir}$ might be expected to shift with $\alpha_{\rm CO}$ accordingly (see Section \ref{sec:virial}), but this shift would likely be in the sense that clouds would be more gravitationally bound than what we measure. 

In Figure \ref{fig:virial_distance} we plot $\alpha_{\rm vir}$ as a function of projected distance from the center of Henize 2-10 for each GMC in our sample.  We find no significant statistical trend with distance, consistent  with the findings of \citet{Johnson_2018}, who found that regions of high-density gas near SSCs do not exhibit preferentially large line widths.  This is interesting, because one might expect that feedback associated with the starburst region would lead to higher levels of turbulence in GMCs and, consequently, larger virial parameters than clouds in a relatively quiescent part of the galaxy.  However, feedback could also lead to increased compression of the gas, resulting in low virial parameters.  Ultimately, we would benefit from more information to determine whether this is a real trend that is probing the underlying physics and chemistry of Henize 2-10.  For instance, if metallicity varies systematically across the galaxy, it might be appropriate to apply a variable CO-to-\htwo~conversion factor to calculate \mlum~for individual clouds.

Observational studies including \citet{Johnson_2000}, \citet{Santangelo_2009}, and \citet{Johnson_2018} have suggested that the molecular clouds in Henize 2-10 may be the sites of future super-star clusters. The fraction $\epsilon$ of gas in a GMC that is converted into stars is roughly $5\%$ in the Milky Way \citep[e.g.,][]{Williams_1997, Evans_2009, Murray_2010}, and it may increase to as high as $\sim 35\%$, as a function of gas surface density, in luminous starburst galaxies \citep{Murray_2010}. \citet{Johnson_2000} estimate a total starburst mass in regions A and B of roughly 1.6-$2.6\times10^6$ \msun~and 2.6-$6.6\times10^4$ \msun, respectively.  Assuming $\epsilon=5\%$, 47 of the clouds in our catalog, including 5 clouds in the southern "tail," currently have enough molecular mass to form SSCs similar in mass to region B.  Of these, 33 clouds---all in the main body of the galaxy---have virial parameters less than unity.

It is also possible that the efficiency of converting gas to stars is greater than $5\%$ in Henize 2-10.   For instance, if the efficiency were as high as $35\%$, Cloud 112 has sufficiently mass at present ($\mlum=5.1\times10^6$ \msun) to eventually form an SSC with mass comparable to region A.   Whatever the alternatives, it is entirely plausible that, with their roughly unity virial parameters, and given reasonable efficiencies, the GMCs in Henize 2-10 may be the precursors of subsequent generations of SSCs.

\section{Summary}\label{sec:summary}
We presented new \co{12} ALMA observations of the blue compact dwarf galaxy, Henize 2-10.  To date, these are the highest resolution observations of molecular gas in this galaxy.  We used these data to map the spatial and kinematic structure of the molecular ISM and explore the properties of giant molecular clouds in Henize 2-10.  We summarize our most salient results:
\begin{enumerate}
    \item The molecular gas of Henize 2-10 has a complex morphology, with clumps, ring-like structures, and high-density peaks of emission.  Most of the gas, about $70\%$, is concentrated in the north of the galaxy and is associated with the optical peak, where most of the starburst activity is taking place.  Assuming a standard CO-to-\htwo~conversion factor for the Milky Way Galaxy, we calculate a total molecular gas mass in the galaxy of $(1.2\pm 0.4)\times 10^8$ \msun.  Roughly $30\%$ of this mass resides to the south-east in an extended, 320-pc long stream of gas.
    
    \item Henize 2-10 has a velocity gradient in the molecular gas within the inner 70 pc around starburst region A, consistent with solid body rotation.  Assuming edge-on rotation, the gradient implies a lower limit to the dynamical mass of $2.7\times10^6$ \msun.  The combined mass of the star clusters and the supermassive black hole candidate in the region is roughly $4\times10^6$ \msun, suggesting that the stars and potential SMBH make the dominant contribution to the dynamical mass, unless the structure has a lower inclination.  Given the total mass within 70 pc of starburst region A, including $6\times 10^6$ \msun~of molecular gas, we constrain the inclination to $31\degr-35\degr$.
    
    \item We used the \texttt{CPROPS} algorithm \citep{Rosolowsky_2006} to identify molecular clouds in the three-dimensional data set.  We identify 178 GMCs, of which 119 are resolved and are used for our final sample.  The clouds have median properties similar to those in the Molecular Ring of the Milky Way, including size ($26$ pc), luminous mass ($4.0\times 10^5$ \msun), and surface density ($180$ \sunits) for Henize 2-10 clouds. Henize 2-10 GMC velocity dispersions (median $3.2$ \kms), are about a factor 50\% higher, at the same mass or size, than those in the Milky Way.
    
    \item Henize 2-10 GMCs have a mass spectrum with a power law slope of $-1.54\pm 0.10$ and a truncation mass of $(4.7\pm0.8)\times 10^6$ \msun, remarkably consistent to that measured for Milky Way clouds.
    
    \item The size-linewidth relation for GMCs in Henize 2-10 shows a large amount of scatter but these properties are formally correlated at the $4\sigma$ level based on the Pearson coefficient.  Overall, the GMCs occupy a similar region of the size-linewidth parameter space as clouds in the disks of the Milky Way and other galaxies, but the formal slope of the size-linewidth relation in Henize 2-10 is 1.3, significantly steeper than that in the comparison samples.
    
    \item The luminous mass versus virial mass relationship for Henize 2-10 GMCs shows a high degree of correlation and a slope of $1.2\pm 0.1$. The Henize 2-10 clouds also lie along the locus of gravitational equilibrium in the size-linewidth coefficient -- surface density plane, though slightly higher in both parameters than disk galaxy clouds to which we compare. This supports the conclusion that the Henize 2-10 clouds are largely in or near virial equilibrium.
    
    \item Assuming a constant CO-to-\htwo~conversion factor, Mass and size are closely correlated in the Henize 2-10 GMCs, with a slope of $3.0\pm 0.3$.  Our results suggests that, in contrast to the Milky Way, instead of constant surface density, Henize 2-10 GMCs have constant volume density.
    
    \item The average molecular gas surface density in Henize 2-10 is a factor of 30 to 70 higher than in the Milky Way disk, despite the fact that their GMC surface densities are similar. This reflects the fact that the molecular gas filling factor in Henize 2-10 is close to unity.
    
    \item The most massive GMCs in Henize 2-10 are potential sites for future super-star clusters because (1) they have near-unity virial parameters, so they are gravitationally bound and thus can form stars; and (2) given a reasonable efficiency of converting gas to stars (as low as $\sim 5\%$) they will form clusters with masses in the range of the existing SSCs in the galaxy.
    
\end{enumerate}

\acknowledgements
We thank the referee for insightful comments and questions which helped to improve this paper.  We also thank Alberto Bolatto for his comments on an early draft of the paper and I-Ting Ho for sharing his valuable expertise on interpreting metallicity observations and measurements.   

This work was supported by the John Harvard Distinguished Science Fellowship at Harvard University.  

This paper makes use of the following ALMA data: ADS/JAO.ALMA\#2015.1.01395.S. ALMA is a partnership of ESO (representing its member states), NSF (USA) and NINS (Japan), together with NRC (Canada) and NSC and ASIAA (Taiwan) and KASI (Republic of Korea), in cooperation with the Republic of Chile. The Joint ALMA Observatory is operated by ESO, AUI/NRAO and NAOJ.

The National Radio Astronomy Observatory is a facility of the National Science Foundation operated under cooperative agreement by Associated Universities, Inc.

\bibliography{paper}

\clearpage

\input{table_cloudprops.tex}
\clearpage

\end{document}

%% file: table_he210.tex
\begin{table}\label{table1}\centering
\begin{center}
\begin{tabular}{lcc}
\multicolumn{3}{c}{Table 1: Properties of Henize 2-10.}\\
\tableline\tableline
Property       & Value      &   Reference \\
\tableline
Distance       & 8.7 Mpc    &  1 \\
Absolute $B$ magnitude  &    $-18.1$     &  2 \\
$12+\log(\rm{O/H})$     & $8.55\pm 0.02$  &  3 \\
Size (major$\times$minor)  & $1\farcm9\times 1\farcm4$  &  4  \\
Stellar mass   & $3.7\times 10^9 \msun$  &  5 \\
Dynamical mass & $(6.3\pm 3.2)\times 10^9~\msun$  & 6  \\
FIR luminosity    &   $4.6\times 10^9~L_\odot$  & 4  \\
SFR$_{\rm FIR}$   & 0.79 \msun~yr$^{-1}$  & 4 \\
SFR$_{{\rm H}\alpha}$    & 1.9 \msun~yr$^{-1}$    &  5 \\
\HI~mass       &    $3.1\times 10^8 \msun$         &  7 \\ 
\htwo~mass     &    $(1.2\pm0.4)\times 10^8~\msun$        & This work  \\
Molecular depletion  time  & $0.15$ Gyr  & This work   \\
\tableline
\end{tabular}
\begin{tablenotes}
\item {\textbf{References}. (1) \citet{Tully_1988}; (2) \citet{Johansson_1987}; (3) \citet{Esteban_2014}; (4) \citet{Madden_2014}; (5) \citet{Reines_2011};  (6) \citet{Baas_1994}; (7) \citet{Sauvage_1997} }
\item \textbf{Note}. The molecular gas depletion time is defined:  $\tau_{\rm dep}\equiv \mhtwo/\rm{SFR}$.
\end{tablenotes}
\end{center}
\end{table}

%% file: table_observations.tex
\begin{table*}[ht]
\centering
\begin{center}
\begin{tabular}{lcc}
\multicolumn{3}{c}{Table 2: Observation Summary.}\label{table2}\\
\tableline\tableline
Date                 & 2016 July 31  &  2016 Aug 2  \\
On-source time       & 31.75 min     &  31.75 min  \\
Number of Antennas   & 36            &  36   \\  
Average $T_{\rm sys}$ & 98.65 K       &  95.94 K     \\
Mean precipitable water vapor & 1.98 mm  &  0.87 mm \\
Bandpass calibrator  & J1037-2934    &  J1037-2934  \\
Flux calibrator      & J1037-2934 &  J1107-4449      \\
Phase calibrator     & J0826-2230 &  J0826-2230      \\
Pointing calibrators & J0826-2230, J1037-2934 &  J0826-2230, J1037-2934   \\
\tableline
\end{tabular}
\end{center}
\end{table*}

%% file: table_cloudmass.tex
\begin{table}\centering
\begin{center}
\begin{tabular}{lcc}
\multicolumn{3}{c}{Table 3: Total luminous mass in \hen~GMCs.}\\
\tableline\tableline
Extrapolation? & $M_{\rm total}$   &  $M_{\rm resolved}$   \\
\tableline
Yes           & $8.8\times 10^7$ \msun & $8.3\times 10^7$ \msun  \\
No            & $5.8\times 10^7$ \msun & $5.5\times 10^7$ \msun  \\
\tableline
\end{tabular}
\begin{tablenotes}
\item \textbf{Notes}. The first column indicates whether the total luminous mass is measured from clouds whose emission has been extrapolated to 0 K. The second column is the total mass in all 178 clouds.  The third column indicates the total mass in the final catalog of 119 resolved clouds.
\end{tablenotes}
\end{center}
\end{table}\label{tab:cloudmass}

%% file: table_cloudprops.tex
\begin{longtable}{ccccccccc}
\multicolumn{9}{c}{Table 4: Properties of Henize 2-10 GMCs.}\\
\hline
Cloud & R.A. & Dec. & $v_0$ & $R$ & $\sigma_v$ & $M_{\rm lum}$ & $M_{\rm vir}$ & $T_{\rm max}$ \\
      & \multicolumn{2}{c}{(J2000)} & (\kms)& (pc)& (\kms) & ($10^5\msun$) & ($10^5\msun$) & (K) \\
\hline
\endfirsthead
\multicolumn{9}{c}
{{\tablename\ \thetable{} -- \emph{continued}}} \\
\hline
Cloud & R.A. & Dec. & $v_0$ & $R$ & $\sigma_v$ & $M_{\rm lum}$ & $M_{\rm vir}$ & $T_{\rm max}$ \\
      & \multicolumn{2}{c}{(J2000)} & (\kms)& (pc)& (\kms) & ($10^5\msun$) & ($10^5\msun$) & (K) \\
      \hline
\endhead
1 & 08:36:15.20 & -26:24:37.5 & 787.6 & $21.6\pm6.6$ & $2.7\pm0.7$ & $1.6\pm0.6$ & $1.6\pm0.9$ & 4.6 \\
2 & 08:36:15.18 & -26:24:37.2 & 794.8 & $12.1\pm3.4$ & $3.7\pm0.7$ & $2.3\pm0.5$ & $1.7\pm0.9$ & 5.0 \\
3 & 08:36:15.94 & -26:24:43.3 & 799.0 & $28.1\pm4.5$ & $2.7\pm0.6$ & $2.9\pm0.6$ & $2.2\pm1.1$ & 5.8 \\
4 & 08:36:15.92 & -26:24:43.1 & 807.8 & $38.1\pm3.6$ & $5.2\pm0.7$ & $9.8\pm0.9$ & $10.6\pm3.2$ & 7.9 \\
5 & 08:36:15.38 & -26:24:32.3 & 806.0 & $18.1\pm6.0$ & $4.8\pm1.1$ & $2.5\pm0.5$ & $4.4\pm2.3$ & 5.2 \\
6 & 08:36:15.59 & -26:24:38.0 & 806.2 & $25.7\pm3.7$ & $4.8\pm0.6$ & $5.6\pm0.6$ & $6.1\pm1.9$ & 7.3 \\
7 & 08:36:15.67 & -26:24:38.6 & 811.7 & $11.0\pm5.5$ & $2.3\pm0.7$ & $1.2\pm0.3$ & $0.6\pm0.4$ & 5.7 \\
8 & 08:36:15.56 & -26:24:37.1 & 814.4 & $36.0\pm3.2$ & $4.5\pm0.5$ & $7.9\pm0.5$ & $7.6\pm1.7$ & 8.1 \\
9 & 08:36:15.66 & -26:24:40.0 & 817.2 & $22.5\pm9.3$ & $2.8\pm1.1$ & $1.4\pm0.5$ & $1.8\pm1.6$ & 4.9 \\
10 & 08:36:15.23 & -26:24:38.2 & 818.3 & $44.0\pm3.7$ & $7.0\pm0.8$ & $12.9\pm0.6$ & $22.4\pm5.9$ & 8.8 \\
11 & 08:36:15.59 & -26:24:35.1 & 825.1 & $38.4\pm3.8$ & $3.1\pm0.5$ & $6.4\pm0.7$ & $3.7\pm1.2$ & 6.1 \\
12 & 08:36:15.59 & -26:24:41.1 & 828.3 & $30.6\pm5.4$ & $3.4\pm0.7$ & $3.9\pm1.0$ & $3.7\pm1.8$ & 4.9 \\
13 & 08:36:15.19 & -26:24:38.2 & 830.2 & $41.2\pm4.1$ & $3.2\pm0.2$ & $8.0\pm0.5$ & $4.3\pm0.8$ & 10.4 \\
14 & 08:36:15.33 & -26:24:31.8 & 823.4 & $43.3\pm2.2$ & $6.4\pm0.4$ & $24.1\pm1.0$ & $18.5\pm2.3$ & 11.0 \\
15 & 08:36:15.30 & -26:24:30.6 & 830.0 & $19.1\pm2.9$ & $4.2\pm0.6$ & $4.2\pm0.4$ & $3.6\pm1.1$ & 7.9 \\
16 & 08:36:15.42 & -26:24:32.3 & 834.0 & $24.8\pm5.5$ & $2.5\pm0.4$ & $2.5\pm0.3$ & $1.6\pm0.7$ & 6.5 \\
17 & 08:36:15.73 & -26:24:41.4 & 835.1 & $15.8\pm6.5$ & $3.5\pm1.1$ & $1.5\pm0.4$ & $2.1\pm1.4$ & 6.0 \\
18 & 08:36:15.26 & -26:24:31.2 & 835.9 & $51.7\pm3.9$ & $3.9\pm0.3$ & $11.4\pm0.6$ & $8.3\pm1.5$ & 11.5 \\
19 & 08:36:15.61 & -26:24:41.3 & 835.5 & $25.3\pm5.9$ & $3.4\pm0.7$ & $2.5\pm1.1$ & $3.0\pm1.5$ & 4.7 \\
20 & 08:36:15.34 & -26:24:37.6 & 831.7 & $31.7\pm2.2$ & $5.1\pm0.4$ & $13.2\pm0.6$ & $8.6\pm1.4$ & 10.8 \\
21 & 08:36:15.57 & -26:24:34.9 & 834.5 & $48.8\pm2.7$ & $4.3\pm0.3$ & $17.9\pm0.7$ & $9.3\pm1.3$ & 8.8 \\
22 & 08:36:15.35 & -26:24:35.1 & 833.4 & $20.5\pm2.6$ & $5.2\pm0.3$ & $9.4\pm0.4$ & $5.7\pm1.0$ & 12.2 \\
23 & 08:36:15.37 & -26:24:31.7 & 838.3 & $39.5\pm4.8$ & $3.5\pm0.5$ & $5.9\pm0.6$ & $5.0\pm1.6$ & 6.4 \\
24 & 08:36:15.60 & -26:24:39.9 & 830.6 & $57.6\pm3.8$ & $7.0\pm0.5$ & $20.4\pm1.0$ & $29.5\pm4.8$ & 8.7 \\
25 & 08:36:15.43 & -26:24:37.5 & 838.8 & $29.6\pm3.2$ & $4.5\pm0.4$ & $7.3\pm0.4$ & $6.3\pm1.3$ & 9.4 \\
26 & 08:36:15.57 & -26:24:37.6 & 828.7 & $49.8\pm1.8$ & $7.7\pm0.3$ & $41.4\pm1.2$ & $31.1\pm3.0$ & 11.1 \\
27 & 08:36:15.28 & -26:24:31.9 & 839.6 & $33.3\pm3.4$ & $4.1\pm0.5$ & $6.8\pm0.4$ & $5.8\pm1.6$ & 10.6 \\
28 & 08:36:15.53 & -26:24:41.0 & 839.5 & $44.8\pm5.2$ & $4.1\pm0.5$ & $7.8\pm1.1$ & $7.8\pm2.2$ & 7.0 \\
29 & 08:36:15.36 & -26:24:37.4 & 841.4 & $19.5\pm2.8$ & $3.5\pm0.5$ & $4.8\pm0.4$ & $2.5\pm0.9$ & 10.6 \\
30 & 08:36:15.32 & -26:24:30.4 & 842.4 & $21.2\pm3.3$ & $3.0\pm0.5$ & $3.3\pm0.6$ & $2.0\pm0.7$ & 6.9 \\
31 & 08:36:15.40 & -26:24:40.1 & 843.1 & $12.4\pm4.3$ & $1.8\pm0.6$ & $1.2\pm0.7$ & $0.4\pm0.3$ & 4.5 \\
32 & 08:36:15.56 & -26:24:37.2 & 843.1 & $25.8\pm3.1$ & $2.1\pm0.2$ & $4.6\pm0.4$ & $1.2\pm0.3$ & 11.0 \\
33 & 08:36:15.41 & -26:24:41.0 & 843.6 & $22.1\pm13.6$ & $2.8\pm1.0$ & $1.1\pm0.6$ & $1.8\pm1.8$ & 4.0 \\
34 & 08:36:15.54 & -26:24:35.7 & 848.4 & $36.9\pm2.1$ & $7.5\pm0.4$ & $22.3\pm1.0$ & $21.4\pm2.4$ & 10.6 \\
35 & 08:36:15.25 & -26:24:32.6 & 843.0 & $44.7\pm3.2$ & $3.1\pm0.3$ & $10.3\pm0.5$ & $4.6\pm0.9$ & 11.3 \\
36 & 08:36:15.36 & -26:24:31.3 & 843.8 & $29.4\pm5.8$ & $3.1\pm0.7$ & $2.8\pm0.6$ & $2.9\pm1.5$ & 6.2 \\
37 & 08:36:15.56 & -26:24:40.2 & 848.8 & $49.4\pm2.5$ & $5.5\pm0.3$ & $22.7\pm1.2$ & $15.4\pm1.8$ & 8.5 \\
38 & 08:36:15.49 & -26:24:43.2 & 844.2 & $10.0\pm7.0$ & $2.5\pm1.1$ & $1.1\pm0.5$ & $0.7\pm0.7$ & 4.2 \\
39 & 08:36:15.18 & -26:24:38.2 & 847.8 & $44.0\pm1.9$ & $7.8\pm0.3$ & $32.8\pm0.9$ & $27.8\pm2.8$ & 11.2 \\
40 & 08:36:15.31 & -26:24:35.1 & 844.5 & $26.7\pm2.7$ & $3.2\pm0.3$ & $7.7\pm0.4$ & $2.8\pm0.7$ & 13.0 \\
41 & 08:36:15.57 & -26:24:38.1 & 844.5 & $37.7\pm4.1$ & $4.1\pm0.3$ & $11.3\pm0.5$ & $6.7\pm1.4$ & 9.7 \\
42 & 08:36:15.18 & -26:24:31.4 & 844.3 & $19.5\pm5.6$ & $2.0\pm0.4$ & $1.4\pm0.4$ & $0.8\pm0.4$ & 5.7 \\
43 & 08:36:15.42 & -26:24:40.7 & 850.4 & $16.3\pm7.5$ & $2.6\pm1.1$ & $1.4\pm0.7$ & $1.1\pm1.2$ & 4.0 \\
44 & 08:36:15.47 & -26:24:34.2 & 846.9 & $31.3\pm4.9$ & $2.3\pm0.4$ & $4.0\pm0.5$ & $1.8\pm0.6$ & 7.6 \\
45 & 08:36:15.18 & -26:24:31.4 & 848.6 & $19.5\pm4.5$ & $1.8\pm0.5$ & $1.2\pm0.3$ & $0.6\pm0.4$ & 6.4 \\
46 & 08:36:15.51 & -26:24:37.5 & 850.0 & $47.3\pm3.4$ & $3.1\pm0.2$ & $11.7\pm0.5$ & $4.7\pm0.9$ & 10.6 \\
47 & 08:36:15.26 & -26:24:32.1 & 849.6 & $49.1\pm4.1$ & $2.1\pm0.2$ & $10.2\pm0.5$ & $2.3\pm0.6$ & 12.1 \\
48 & 08:36:15.34 & -26:24:30.4 & 852.4 & $18.5\pm8.1$ & $4.4\pm1.2$ & $2.2\pm0.8$ & $3.7\pm2.6$ & 5.0 \\
49 & 08:36:15.43 & -26:24:39.4 & 849.5 & $14.4\pm5.3$ & $1.7\pm0.6$ & $1.4\pm0.5$ & $0.5\pm0.4$ & 4.6 \\
50 & 08:36:15.09 & -26:24:37.7 & 850.2 & $44.9\pm3.2$ & $5.6\pm0.4$ & $15.4\pm0.8$ & $14.7\pm2.2$ & 8.6 \\
51 & 08:36:15.32 & -26:24:35.2 & 863.0 & $37.3\pm2.4$ & $8.5\pm0.4$ & $26.5\pm0.7$ & $27.8\pm2.8$ & 12.9 \\
52 & 08:36:15.48 & -26:24:34.4 & 852.2 & $18.3\pm5.0$ & $2.4\pm0.6$ & $1.9\pm0.3$ & $1.1\pm0.6$ & 7.4 \\
53 & 08:36:15.15 & -26:24:30.5 & 851.8 & $29.7\pm4.1$ & $5.0\pm0.7$ & $6.0\pm0.5$ & $7.7\pm2.8$ & 6.9 \\
54 & 08:36:15.51 & -26:24:43.3 & 855.8 & $37.1\pm5.8$ & $5.2\pm0.8$ & $7.6\pm1.4$ & $10.5\pm3.2$ & 6.3 \\
55 & 08:36:15.38 & -26:24:39.6 & 853.6 & $17.1\pm6.0$ & $2.0\pm0.6$ & $1.4\pm0.3$ & $0.7\pm0.5$ & 6.0 \\
56 & 08:36:15.55 & -26:24:39.4 & 854.9 & $31.7\pm5.6$ & $2.1\pm0.4$ & $2.4\pm0.5$ & $1.5\pm0.7$ & 6.8 \\
57 & 08:36:15.41 & -26:24:34.2 & 853.6 & $29.9\pm4.3$ & $2.7\pm0.4$ & $3.7\pm0.4$ & $2.3\pm0.7$ & 6.6 \\
58 & 08:36:15.53 & -26:24:42.0 & 862.4 & $56.0\pm4.4$ & $8.6\pm0.7$ & $15.4\pm1.2$ & $43.5\pm9.1$ & 6.3 \\
59 & 08:36:15.39 & -26:24:39.9 & 855.7 & $12.8\pm8.4$ & $1.1\pm0.7$ & $0.4\pm0.3$ & $0.2\pm0.2$ & 5.4 \\
60 & 08:36:15.53 & -26:24:37.6 & 856.8 & $45.5\pm3.3$ & $2.2\pm0.2$ & $8.5\pm0.7$ & $2.4\pm0.5$ & 8.5 \\
61 & 08:36:15.13 & -26:24:34.4 & 853.8 & $23.4\pm2.8$ & $4.0\pm0.4$ & $6.6\pm0.4$ & $3.9\pm1.0$ & 8.0 \\
62 & 08:36:15.25 & -26:24:32.8 & 856.3 & $54.0\pm2.5$ & $4.0\pm0.3$ & $19.9\pm0.7$ & $8.8\pm1.2$ & 13.9 \\
63 & 08:36:15.37 & -26:24:39.7 & 859.7 & $16.7\pm6.5$ & $2.1\pm0.6$ & $1.1\pm0.3$ & $0.8\pm0.6$ & 5.9 \\
64 & 08:36:15.40 & -26:24:34.2 & 859.0 & $30.5\pm4.5$ & $1.3\pm0.2$ & $3.1\pm0.3$ & $0.5\pm0.1$ & 8.3 \\
65 & 08:36:15.37 & -26:24:42.7 & 863.7 & $22.0\pm10.6$ & $2.9\pm1.7$ & $1.1\pm0.5$ & $1.9\pm2.4$ & 4.5 \\
66 & 08:36:15.34 & -26:24:40.9 & 862.0 & $13.6\pm8.4$ & $2.8\pm1.0$ & $1.4\pm0.4$ & $1.1\pm1.2$ & 5.1 \\
67 & 08:36:15.02 & -26:24:39.1 & 861.4 & $21.2\pm7.0$ & $3.5\pm0.8$ & $1.8\pm0.6$ & $2.7\pm1.5$ & 5.6 \\
68 & 08:36:15.08 & -26:24:38.7 & 862.6 & $42.1\pm5.9$ & $3.0\pm0.3$ & $5.2\pm0.7$ & $3.9\pm1.1$ & 7.9 \\
69 & 08:36:15.42 & -26:24:35.4 & 859.7 & $13.2\pm9.4$ & $1.5\pm0.9$ & $0.7\pm0.4$ & $0.3\pm0.4$ & 4.0 \\
70 & 08:36:15.51 & -26:24:37.9 & 864.9 & $36.9\pm3.9$ & $3.5\pm0.4$ & $7.0\pm0.7$ & $4.6\pm1.0$ & 6.5 \\
71 & 08:36:15.48 & -26:24:35.2 & 863.8 & $17.7\pm3.9$ & $3.9\pm0.6$ & $3.7\pm0.4$ & $2.8\pm1.0$ & 7.0 \\
72 & 08:36:15.10 & -26:24:34.4 & 865.4 & $35.4\pm2.8$ & $5.2\pm0.3$ & $14.1\pm0.6$ & $9.9\pm1.5$ & 10.6 \\
73 & 08:36:15.40 & -26:24:34.2 & 865.6 & $30.6\pm3.5$ & $2.7\pm0.3$ & $6.0\pm0.4$ & $2.4\pm0.6$ & 8.8 \\
74 & 08:36:15.52 & -26:24:40.1 & 865.1 & $49.5\pm4.7$ & $3.7\pm0.3$ & $11.3\pm0.9$ & $7.2\pm1.4$ & 7.9 \\
75 & 08:36:15.50 & -26:24:36.5 & 864.6 & $30.4\pm3.8$ & $3.1\pm0.5$ & $4.2\pm0.5$ & $3.1\pm1.0$ & 6.9 \\
76 & 08:36:15.22 & -26:24:32.4 & 865.0 & $29.7\pm3.3$ & $2.4\pm0.2$ & $7.1\pm0.4$ & $1.8\pm0.4$ & 13.5 \\
77 & 08:36:15.01 & -26:24:39.1 & 869.2 & $31.6\pm7.9$ & $2.6\pm0.5$ & $2.4\pm0.7$ & $2.1\pm1.0$ & 4.5 \\
78 & 08:36:15.10 & -26:24:38.8 & 869.0 & $19.5\pm5.7$ & $3.1\pm1.1$ & $1.7\pm0.4$ & $2.0\pm1.5$ & 6.1 \\
79 & 08:36:15.20 & -26:24:32.4 & 873.5 & $23.0\pm2.8$ & $4.8\pm0.5$ & $8.1\pm0.4$ & $5.5\pm1.5$ & 12.6 \\
80 & 08:36:15.31 & -26:24:40.9 & 874.1 & $31.5\pm5.9$ & $4.4\pm0.7$ & $4.2\pm0.8$ & $6.2\pm2.5$ & 5.1 \\
81 & 08:36:15.44 & -26:24:41.2 & 871.3 & $24.4\pm7.9$ & $2.1\pm0.6$ & $1.7\pm0.5$ & $1.1\pm0.8$ & 4.7 \\
82 & 08:36:15.49 & -26:24:36.6 & 870.5 & $22.6\pm6.3$ & $1.2\pm0.3$ & $1.6\pm0.3$ & $0.3\pm0.2$ & 6.8 \\
83 & 08:36:15.50 & -26:24:39.9 & 873.0 & $44.9\pm5.6$ & $2.8\pm0.4$ & $5.2\pm0.6$ & $3.6\pm1.1$ & 7.0 \\
84 & 08:36:15.39 & -26:24:40.1 & 873.3 & $5.3\pm5.8$ & $1.0\pm0.7$ & $0.4\pm0.2$ & $0.1\pm0.1$ & 4.7 \\
85 & 08:36:15.50 & -26:24:38.3 & 875.4 & $34.1\pm5.7$ & $2.5\pm0.5$ & $3.2\pm0.4$ & $2.2\pm0.8$ & 6.9 \\
86 & 08:36:15.48 & -26:24:36.6 & 876.7 & $15.7\pm7.7$ & $3.1\pm1.1$ & $1.4\pm0.3$ & $1.6\pm1.5$ & 5.4 \\
87 & 08:36:15.00 & -26:24:32.1 & 875.2 & $20.0\pm8.3$ & $1.3\pm0.7$ & $1.0\pm0.3$ & $0.3\pm0.4$ & 4.7 \\
88 & 08:36:15.51 & -26:24:39.6 & 878.5 & $35.9\pm5.3$ & $3.9\pm0.6$ & $5.0\pm0.6$ & $5.6\pm1.8$ & 6.1 \\
89 & 08:36:15.00 & -26:24:32.4 & 878.2 & $19.8\pm8.0$ & $1.5\pm0.7$ & $0.9\pm0.2$ & $0.5\pm0.5$ & 5.5 \\
90 & 08:36:15.01 & -26:24:34.2 & 876.9 & $39.2\pm2.9$ & $3.9\pm0.3$ & $10.6\pm0.5$ & $6.1\pm1.1$ & 10.6 \\
91 & 08:36:15.44 & -26:24:41.4 & 882.8 & $14.6\pm7.4$ & $2.3\pm1.3$ & $1.1\pm0.5$ & $0.8\pm1.0$ & 3.9 \\
92 & 08:36:15.42 & -26:24:40.2 & 880.6 & $9.0\pm4.8$ & $3.5\pm0.6$ & $1.5\pm0.5$ & $1.1\pm0.7$ & 4.2 \\
93 & 08:36:15.51 & -26:24:38.2 & 881.2 & $14.5\pm6.7$ & $1.7\pm0.6$ & $1.0\pm0.3$ & $0.4\pm0.4$ & 5.5 \\
94 & 08:36:15.05 & -26:24:34.1 & 881.8 & $20.3\pm3.3$ & $2.6\pm0.4$ & $3.2\pm0.3$ & $1.4\pm0.5$ & 9.3 \\
95 & 08:36:15.09 & -26:24:35.2 & 882.2 & $26.3\pm3.7$ & $3.2\pm0.5$ & $3.7\pm0.5$ & $2.8\pm0.9$ & 6.0 \\
96 & 08:36:15.43 & -26:24:38.7 & 888.3 & $38.2\pm3.7$ & $4.4\pm0.6$ & $7.1\pm0.7$ & $7.7\pm2.2$ & 6.4 \\
97 & 08:36:15.08 & -26:24:35.2 & 892.7 & $19.9\pm5.8$ & $4.5\pm1.0$ & $2.7\pm0.6$ & $4.3\pm2.2$ & 4.9 \\
98 & 08:36:15.04 & -26:24:33.9 & 890.5 & $17.2\pm2.5$ & $3.8\pm0.4$ & $5.6\pm0.4$ & $2.6\pm0.6$ & 9.3 \\
99 & 08:36:14.99 & -26:24:33.6 & 889.3 & $24.3\pm4.1$ & $3.3\pm0.4$ & $4.7\pm0.4$ & $2.8\pm0.9$ & 10.5 \\
100 & 08:36:14.99 & -26:24:32.6 & 887.7 & $11.9\pm5.1$ & $1.8\pm0.5$ & $1.3\pm0.3$ & $0.4\pm0.3$ & 7.8 \\
101 & 08:36:15.40 & -26:24:40.1 & 890.4 & $29.2\pm10.3$ & $3.0\pm0.7$ & $2.2\pm0.7$ & $2.8\pm1.4$ & 4.8 \\
102 & 08:36:15.44 & -26:24:38.5 & 897.3 & $40.0\pm5.2$ & $2.1\pm0.3$ & $4.6\pm0.4$ & $1.8\pm0.6$ & 6.6 \\
103 & 08:36:15.05 & -26:24:33.7 & 901.6 & $14.4\pm2.2$ & $4.3\pm0.6$ & $4.4\pm0.4$ & $2.8\pm0.9$ & 9.0 \\
104 & 08:36:15.44 & -26:24:39.7 & 900.8 & $16.8\pm9.7$ & $2.8\pm1.1$ & $1.1\pm0.4$ & $1.4\pm1.3$ & 4.4 \\
105 & 08:36:14.99 & -26:24:33.3 & 900.8 & $20.0\pm2.1$ & $6.7\pm0.7$ & $9.2\pm0.4$ & $9.4\pm2.2$ & 10.0 \\
106 & 08:36:15.43 & -26:24:38.6 & 902.8 & $33.4\pm4.9$ & $2.6\pm0.4$ & $2.8\pm0.5$ & $2.4\pm0.9$ & 5.9 \\
107 & 08:36:15.45 & -26:24:38.5 & 906.1 & $23.9\pm4.5$ & $3.0\pm0.5$ & $3.0\pm0.4$ & $2.2\pm0.8$ & 7.4 \\
108 & 08:36:15.58 & -26:24:51.7 & 825.2 & $76.7\pm4.5$ & $7.8\pm0.7$ & $30.0\pm3.2$ & $48.2\pm9.3$ & 7.3 \\
109 & 08:36:15.67 & -26:24:54.8 & 829.3 & $13.6\pm11.2$ & $2.7\pm1.3$ & $1.4\pm0.8$ & $1.0\pm1.4$ & 7.8 \\
110 & 08:36:15.73 & -26:24:52.5 & 822.5 & $69.4\pm4.9$ & $6.0\pm0.6$ & $29.8\pm3.2$ & $25.8\pm6.2$ & 8.4 \\
111 & 08:36:15.87 & -26:24:49.5 & 831.2 & $28.4\pm11.8$ & $1.6\pm0.8$ & $1.9\pm0.6$ & $0.8\pm0.9$ & 6.3 \\
112 & 08:36:15.88 & -26:24:48.6 & 823.9 & $72.3\pm3.0$ & $10.1\pm0.6$ & $51.1\pm2.9$ & $76.0\pm8.9$ & 9.7 \\
113 & 08:36:15.74 & -26:24:48.1 & 836.7 & $10.4\pm5.5$ & $3.2\pm1.5$ & $1.5\pm0.8$ & $1.1\pm1.4$ & 4.7 \\
114 & 08:36:15.65 & -26:24:49.9 & 843.8 & $43.0\pm6.5$ & $6.9\pm1.1$ & $10.2\pm1.9$ & $21.4\pm7.6$ & 8.0 \\
115 & 08:36:15.09 & -26:24:36.2 & 817.9 & $17.2\pm7.2$ & $6.3\pm1.9$ & $2.4\pm0.7$ & $7.1\pm5.1$ & 4.5 \\
116 & 08:36:14.68 & -26:24:36.6 & 860.7 & $13.9\pm5.1$ & $4.7\pm0.9$ & $2.8\pm0.8$ & $3.1\pm1.6$ & 4.7 \\
117 & 08:36:14.70 & -26:24:35.3 & 878.2 & $14.6\pm3.0$ & $5.0\pm0.6$ & $4.2\pm0.4$ & $3.8\pm1.3$ & 7.4 \\
118 & 08:36:14.42 & -26:24:24.7 & 880.1 & $19.2\pm5.2$ & $4.4\pm1.2$ & $3.1\pm0.7$ & $3.8\pm2.1$ & 6.4 \\
119 & 08:36:14.47 & -26:24:34.1 & 887.2 & $22.9\pm5.2$ & $4.2\pm1.1$ & $3.3\pm0.8$ & $4.3\pm2.3$ & 4.6 \\
\hline
\end{longtable}\label{tab:cloudprops} 

%% file: paper.bbl
\begin{thebibliography}{}
\expandafter\ifx\csname natexlab\endcsname\relax\def\natexlab#1{#1}\fi

\bibitem[{{Asplund} {et~al.}(2009){Asplund}, {Grevesse}, {Sauval}, \&
  {Scott}}]{Asplund_2009}
{Asplund}, M., {Grevesse}, N., {Sauval}, A.~J., \& {Scott}, P. 2009, \araa, 47,
  481

\bibitem[{{Baas} {et~al.}(1994){Baas}, {Israel}, \& {Koornneef}}]{Baas_1994}
{Baas}, F., {Israel}, F.~P., \& {Koornneef}, J. 1994, \aap, 284, 403

\bibitem[{{Bacon} {et~al.}(2010){Bacon}, {Accardo}, {Adjali}, {Anwand},
  {Bauer}, {Biswas}, {Blaizot}, {Boudon}, {Brau-Nogue}, {Brinchmann},
  {Caillier}, {Capoani}, {Carollo}, {Contini}, {Couderc}, {Daguis{\'e}},
  {Deiries}, {Delabre}, {Dreizler}, {Dubois}, {Dupieux}, {Dupuy}, {Emsellem},
  {Fechner}, {Fleischmann}, {Fran{\c c}ois}, {Gallou}, {Gharsa}, {Glindemann},
  {Gojak}, {Guiderdoni}, {Hansali}, {Hahn}, {Jarno}, {Kelz}, {Koehler},
  {Kosmalski}, {Laurent}, {Le Floch}, {Lilly}, {Lizon}, {Loupias}, {Manescau},
  {Monstein}, {Nicklas}, {Olaya}, {Pares}, {Pasquini}, {P{\'e}contal-Rousset},
  {Pell{\'o}}, {Petit}, {Popow}, {Reiss}, {Remillieux}, {Renault}, {Roth},
  {Rupprecht}, {Serre}, {Schaye}, {Soucail}, {Steinmetz}, {Streicher}, {Stuik},
  {Valentin}, {Vernet}, {Weilbacher}, {Wisotzki}, \& {Yerle}}]{Bacon_2010}
{Bacon}, R., {Accardo}, M., {Adjali}, L., {et~al.} 2010, in \procspie, Vol.
  7735, Ground-based and Airborne Instrumentation for Astronomy III, 773508

\bibitem[{{Bayet} {et~al.}(2004){Bayet}, {Gerin}, {Phillips}, \&
  {Contursi}}]{Bayet_2004}
{Bayet}, E., {Gerin}, M., {Phillips}, T.~G., \& {Contursi}, A. 2004, \aap, 427,
  45

\bibitem[{{Bertoldi} \& {McKee}(1992)}]{Bertoldi_1992}
{Bertoldi}, F., \& {McKee}, C.~F. 1992, \apj, 395, 140

\bibitem[{{Bolatto} {et~al.}(2008){Bolatto}, {Leroy}, {Rosolowsky}, {Walter},
  \& {Blitz}}]{Bolatto_2008}
{Bolatto}, A.~D., {Leroy}, A.~K., {Rosolowsky}, E., {Walter}, F., \& {Blitz},
  L. 2008, \apj, 686, 948

\bibitem[{{Bolatto} {et~al.}(2013){Bolatto}, {Wolfire}, \&
  {Leroy}}]{Bolatto_2013}
{Bolatto}, A.~D., {Wolfire}, M., \& {Leroy}, A.~K. 2013, \araa, 51, 207

\bibitem[{{Colombo} {et~al.}(2014){Colombo}, {Hughes}, {Schinnerer}, {Meidt},
  {Leroy}, {Pety}, {Dobbs}, {Garc{\'{\i}}a-Burillo}, {Dumas}, {Thompson},
  {Schuster}, \& {Kramer}}]{Colombo_2014}
{Colombo}, D., {Hughes}, A., {Schinnerer}, E., {et~al.} 2014, \apj, 784, 3

\bibitem[{{Cresci} {et~al.}(2017){Cresci}, {Vanzi}, {Telles}, {Lanzuisi},
  {Brusa}, {Mingozzi}, {Sauvage}, \& {Johnson}}]{Cresci_2017}
{Cresci}, G., {Vanzi}, L., {Telles}, E., {et~al.} 2017, \aap, 604, A101

\bibitem[{{Dame} {et~al.}(1986){Dame}, {Elmegreen}, {Cohen}, \&
  {Thaddeus}}]{Dame_1986}
{Dame}, T.~M., {Elmegreen}, B.~G., {Cohen}, R.~S., \& {Thaddeus}, P. 1986,
  \apj, 305, 892

\bibitem[{{Dame} {et~al.}(2001){Dame}, {Hartmann}, \& {Thaddeus}}]{Dame_2001}
{Dame}, T.~M., {Hartmann}, D., \& {Thaddeus}, P. 2001, \apj, 547, 792

\bibitem[{{Deharveng} {et~al.}(1988){Deharveng}, {Caplan}, {Lequeux},
  {Azzopardi}, {Breysacher}, {Tarenghi}, \& {Westerlund}}]{Deharveng_1988}
{Deharveng}, L., {Caplan}, J., {Lequeux}, J., {et~al.} 1988, \aaps, 73, 407

\bibitem[{{Downes} \& {Solomon}(1998)}]{Downes_1998}
{Downes}, D., \& {Solomon}, P.~M. 1998, \apj, 507, 615

\bibitem[{{Esteban} {et~al.}(2014){Esteban}, {Garc{\'{\i}}a-Rojas}, {Carigi},
  {Peimbert}, {Bresolin}, {L{\'o}pez-S{\'a}nchez}, \&
  {Mesa-Delgado}}]{Esteban_2014}
{Esteban}, C., {Garc{\'{\i}}a-Rojas}, J., {Carigi}, L., {et~al.} 2014, \mnras,
  443, 624

\bibitem[{{Evans} {et~al.}(2010){Evans}, {Primini}, {Glotfelty}, {Anderson},
  {Bonaventura}, {Chen}, {Davis}, {Doe}, {Evans}, {Fabbiano}, {Galle}, {Gibbs},
  {Grier}, {Hain}, {Hall}, {Harbo}, {He}, {Houck}, {Karovska}, {Kashyap},
  {Lauer}, {McCollough}, {McDowell}, {Miller}, {Mitschang}, {Morgan},
  {Mossman}, {Nichols}, {Nowak}, {Plummer}, {Refsdal}, {Rots}, {Siemiginowska},
  {Sundheim}, {Tibbetts}, {Van Stone}, {Winkelman}, \& {Zografou}}]{Evans_2010}
{Evans}, I.~N., {Primini}, F.~A., {Glotfelty}, K.~J., {et~al.} 2010, \apjs,
  189, 37

\bibitem[{{Evans} {et~al.}(2009){Evans}, {Dunham}, {J{\o}rgensen}, {Enoch},
  {Mer{\'{\i}}n}, {van Dishoeck}, {Alcal{\'a}}, {Myers}, {Stapelfeldt},
  {Huard}, {Allen}, {Harvey}, {van Kempen}, {Blake}, {Koerner}, {Mundy},
  {Padgett}, \& {Sargent}}]{Evans_2009}
{Evans}, II, N.~J., {Dunham}, M.~M., {J{\o}rgensen}, J.~K., {et~al.} 2009,
  \apjs, 181, 321

\bibitem[{{Faesi} {et~al.}(2018){Faesi}, {Lada}, \& {Forbrich}}]{Faesi_2018}
{Faesi}, C.~M., {Lada}, C.~J., \& {Forbrich}, J. 2018, \apj, 857, 19

\bibitem[{{Fanelli} {et~al.}(1988){Fanelli}, {O'Connell}, \&
  {Thuan}}]{Fanelli_1988}
{Fanelli}, M.~N., {O'Connell}, R.~W., \& {Thuan}, T.~X. 1988, \apj, 334, 665

\bibitem[{{Field} {et~al.}(2011){Field}, {Blackman}, \& {Keto}}]{Field_2011}
{Field}, G.~B., {Blackman}, E.~G., \& {Keto}, E.~R. 2011, \mnras, 416, 710

\bibitem[{{Gratier} {et~al.}(2012){Gratier}, {Braine}, {Rodriguez-Fernandez},
  {Schuster}, {Kramer}, {Corbelli}, {Combes}, {Brouillet}, {van der Werf}, \&
  {R{\"o}llig}}]{Gratier_2012}
{Gratier}, P., {Braine}, J., {Rodriguez-Fernandez}, N.~J., {et~al.} 2012, \aap,
  542, A108

\bibitem[{{Guseva} {et~al.}(2000){Guseva}, {Izotov}, \& {Thuan}}]{Guseva_2000}
{Guseva}, N.~G., {Izotov}, Y.~I., \& {Thuan}, T.~X. 2000, \apj, 531, 776

\bibitem[{{Harris}(2003)}]{Harris_2003}
{Harris}, W.~E. 2003, in A Decade of Hubble Space Telescope Science, ed.
  M.~{Livio}, K.~{Noll}, \& M.~{Stiavelli}, Vol.~14, 78--100

\bibitem[{{Henize}(1967)}]{Henize_1967}
{Henize}, K.~G. 1967, \apjs, 14, 125

\bibitem[{{Heyer} \& {Dame}(2015)}]{Heyer_2015}
{Heyer}, M., \& {Dame}, T.~M. 2015, \araa, 53, 583

\bibitem[{{Heyer} {et~al.}(2009){Heyer}, {Krawczyk}, {Duval}, \&
  {Jackson}}]{Heyer_2009}
{Heyer}, M., {Krawczyk}, C., {Duval}, J., \& {Jackson}, J.~M. 2009, \apj, 699,
  1092

\bibitem[{{Heyer} {et~al.}(2001){Heyer}, {Carpenter}, \& {Snell}}]{Heyer_2001}
{Heyer}, M.~H., {Carpenter}, J.~M., \& {Snell}, R.~L. 2001, \apj, 551, 852

\bibitem[{{Hughes} {et~al.}(2013){Hughes}, {Meidt}, {Schinnerer}, {Colombo},
  {Pety}, {Leroy}, {Dobbs}, {Garc{\'{\i}}a-Burillo}, {Thompson}, {Dumas},
  {Schuster}, \& {Kramer}}]{Hughes_2013}
{Hughes}, A., {Meidt}, S.~E., {Schinnerer}, E., {et~al.} 2013, \apj, 779, 44

\bibitem[{{Imanishi} {et~al.}(2007){Imanishi}, {Nakanishi}, {Tamura}, {Oi}, \&
  {Kohno}}]{Imanishi_2007}
{Imanishi}, M., {Nakanishi}, K., {Tamura}, Y., {Oi}, N., \& {Kohno}, K. 2007,
  \aj, 134, 2366

\bibitem[{{Imara} \& {Blitz}(2011)}]{Imara_2011a}
{Imara}, N., \& {Blitz}, L. 2011, \apj, 732, 78

\bibitem[{Inutsuka {et~al.}(2015)Inutsuka, Inoue, Iwasaki, \&
  Hosokawa}]{Inutsuka_2015}
Inutsuka, S.-i., Inoue, T., Iwasaki, K., \& Hosokawa, T. 2015, A{\&}A, 580, A49

\bibitem[{{Johansson}(1987)}]{Johansson_1987}
{Johansson}, I. 1987, \aap, 182, 179

\bibitem[{{Johnson} {et~al.}(2018){Johnson}, {Brogan}, {Indebetouw}, {Testi},
  {Wilner}, {Reines}, {Chen}, \& {Vanzi}}]{Johnson_2018}
{Johnson}, K.~E., {Brogan}, C.~L., {Indebetouw}, R., {et~al.} 2018, \apj, 853,
  125

\bibitem[{{Johnson} {et~al.}(2000){Johnson}, {Leitherer}, {Vacca}, \&
  {Conti}}]{Johnson_2000}
{Johnson}, K.~E., {Leitherer}, C., {Vacca}, W.~D., \& {Conti}, P.~S. 2000, \aj,
  120, 1273

\bibitem[{{Kalberla} \& {Kerp}(2009)}]{Kalberla_2009}
{Kalberla}, P.~M.~W., \& {Kerp}, J. 2009, \araa, 47, 27

\bibitem[{Kauffmann {et~al.}(2017)Kauffmann, Pillai, Zhang, Menten, Goldsmith,
  Lu, \& Guzm{\'a}n}]{Kauffmann_2017}
Kauffmann, J., Pillai, T., Zhang, Q., {et~al.} 2017, A{\&}A, 603, A89

\bibitem[{{Kennicutt} \& {Evans}(2012)}]{Kennicutt_2012}
{Kennicutt}, R.~C., \& {Evans}, N.~J. 2012, \araa, 50, 531

\bibitem[{{Kepley} {et~al.}(2016){Kepley}, {Leroy}, {Johnson}, {Sandstrom}, \&
  {Chen}}]{Kepley_2016}
{Kepley}, A.~A., {Leroy}, A.~K., {Johnson}, K.~E., {Sandstrom}, K., \& {Chen},
  C.-H.~R. 2016, \apj, 828, 50

\bibitem[{{Kerr} \& {Lynden-Bell}(1986)}]{Kerr_1986}
{Kerr}, F.~J., \& {Lynden-Bell}, D. 1986, \mnras, 221, 1023

\bibitem[{{Kewley} \& {Ellison}(2008)}]{Kewley_2008}
{Kewley}, L.~J., \& {Ellison}, S.~L. 2008, \apj, 681, 1183

\bibitem[{{Kobulnicky} {et~al.}(1995){Kobulnicky}, {Dickey}, {Sargent}, {Hogg},
  \& {Conti}}]{Kobulnicky_1995}
{Kobulnicky}, H.~A., {Dickey}, J.~M., {Sargent}, A.~I., {Hogg}, D.~E., \&
  {Conti}, P.~S. 1995, \aj, 110, 116

\bibitem[{{Kobulnicky} \& {Johnson}(1999)}]{Kobulnicky_Johnson_1999}
{Kobulnicky}, H.~A., \& {Johnson}, K.~E. 1999, \apj, 527, 154

\bibitem[{{Kobulnicky} {et~al.}(1999){Kobulnicky}, {Kennicutt}, \&
  {Pizagno}}]{Kobulnicky_1999}
{Kobulnicky}, H.~A., {Kennicutt}, Jr., R.~C., \& {Pizagno}, J.~L. 1999, \apj,
  514, 544

\bibitem[{{Kondrat'eva} \& {Kondratjeva}(1972)}]{Kondratjeva_1972}
{Kondrat'eva}, L.~N., \& {Kondratjeva}, L.~N. 1972, Astronomicheskij
  Tsirkulyar, 683, 7

\bibitem[{{Kormendy} \& {Ho}(2013)}]{Kormendy_2013}
{Kormendy}, J., \& {Ho}, L.~C. 2013, \araa, 51, 511

\bibitem[{{Kunth} {et~al.}(1988){Kunth}, {Maurogordato}, \&
  {Vigroux}}]{Kunth_1988}
{Kunth}, D., {Maurogordato}, S., \& {Vigroux}, L. 1988, \aap, 204, 10

\bibitem[{{Lada} {et~al.}(2012){Lada}, {Forbrich}, {Lombardi}, \&
  {Alves}}]{Lada_2012}
{Lada}, C.~J., {Forbrich}, J., {Lombardi}, M., \& {Alves}, J.~F. 2012, \apj,
  745, 190

\bibitem[{{Lada} {et~al.}(2010){Lada}, {Lombardi}, \& {Alves}}]{Lada_2010}
{Lada}, C.~J., {Lombardi}, M., \& {Alves}, J.~F. 2010, \apj, 724, 687

\bibitem[{{Larson}(1981)}]{Larson_1981}
{Larson}, R.~B. 1981, \mnras, 194, 809

\bibitem[{{Leroy} {et~al.}(2006){Leroy}, {Bolatto}, {Walter}, \&
  {Blitz}}]{Leroy_2006}
{Leroy}, A., {Bolatto}, A., {Walter}, F., \& {Blitz}, L. 2006, \apj, 643, 825

\bibitem[{{Leroy} {et~al.}(2015){Leroy}, {Bolatto}, {Ostriker}, {Rosolowsky},
  {Walter}, {Warren}, {Donovan Meyer}, {Hodge}, {Meier}, {Ott}, {Sandstrom},
  {Schruba}, {Veilleux}, \& {Zwaan}}]{Leroy_2015}
{Leroy}, A.~K., {Bolatto}, A.~D., {Ostriker}, E.~C., {et~al.} 2015, \apj, 801,
  25

\bibitem[{{Licquia} \& {Newman}(2015)}]{Licquia_2015}
{Licquia}, T.~C., \& {Newman}, J.~A. 2015, \apj, 806, 96

\bibitem[{{Lombardi} {et~al.}(2010){Lombardi}, {Alves}, \&
  {Lada}}]{Lombardi_2010}
{Lombardi}, M., {Alves}, J., \& {Lada}, C.~J. 2010, \aap, 519, L7

\bibitem[{{Madden} {et~al.}(2013){Madden}, {R{\'e}my-Ruyer}, {Galametz},
  {Cormier}, {Lebouteiller}, {Galliano}, {Hony}, {Bendo}, {Smith}, {Pohlen},
  {Roussel}, {Sauvage}, {Wu}, {Sturm}, {Poglitsch}, {Contursi}, {Doublier},
  {Baes}, {Barlow}, {Boselli}, {Boquien}, {Carlson}, {Ciesla}, {Cooray},
  {Cortese}, {de Looze}, {Irwin}, {Isaak}, {Kamenetzky}, {Karczewski}, {Lu},
  {MacHattie}, {O'Halloran}, {Parkin}, {Rangwala}, {Schirm}, {Schulz},
  {Spinoglio}, {Vaccari}, {Wilson}, \& {Wozniak}}]{Madden_2013}
{Madden}, S.~C., {R{\'e}my-Ruyer}, A., {Galametz}, M., {et~al.} 2013, \pasp,
  125, 600

\bibitem[{{Madden} {et~al.}(2014){Madden}, {R{\'e}my-Ruyer}, {Galametz},
  {Cormier}, {Lebouteiller}, {Galliano}, {Hony}, {Bendo}, {Smith}, {Pohlen},
  {Roussel}, {Sauvage}, {Wu}, {Sturm}, {Poglitsch}, {Contursi}, {Doublier},
  {Baes}, {Barlow}, {Boselli}, {Boquien}, {Carlson}, {Ciesla}, {Cooray},
  {Cortese}, {De Looze}, {Irwin}, {Isaak}, {Kamenetzky}, {Karczewski}, {Lu},
  {MacHattie}, {O'Halloran}, {Parkin}, {Rangwala}, {Schirm}, {Schulz},
  {Spinoglio}, {Vaccari}, {Wilson}, \& {Wozniak}}]{Madden_2014}
---. 2014, \pasp, 126, 1079

\bibitem[{{Magrini} \& {Gon{\c c}alves}(2009)}]{Magrini_2009}
{Magrini}, L., \& {Gon{\c c}alves}, D.~R. 2009, \mnras, 398, 280

\bibitem[{{Mateo}(1998)}]{Mateo_1998}
{Mateo}, M.~L. 1998, \araa, 36, 435

\bibitem[{{Meier} {et~al.}(2001){Meier}, {Turner}, {Crosthwaite}, \&
  {Beck}}]{Meier_2001}
{Meier}, D.~S., {Turner}, J.~L., {Crosthwaite}, L.~P., \& {Beck}, S.~C. 2001,
  \aj, 121, 740

\bibitem[{{Murray} {et~al.}(2010){Murray}, {Quataert}, \&
  {Thompson}}]{Murray_2010}
{Murray}, N., {Quataert}, E., \& {Thompson}, T.~A. 2010, \apj, 709, 191

\bibitem[{{Narayanan} {et~al.}(2011){Narayanan}, {Krumholz}, {Ostriker}, \&
  {Hernquist}}]{Narayanan_2011}
{Narayanan}, D., {Krumholz}, M., {Ostriker}, E.~C., \& {Hernquist}, L. 2011,
  \mnras, 418, 664

\bibitem[{{Nguyen} {et~al.}(2014){Nguyen}, {Seth}, {Reines}, {den Brok},
  {Sand}, \& {McLeod}}]{Nguyen_2014}
{Nguyen}, D.~D., {Seth}, A.~C., {Reines}, A.~E., {et~al.} 2014, \apj, 794, 34

\bibitem[{{Oka} {et~al.}(2001){Oka}, {Hasegawa}, {Sato}, {Tsuboi}, {Miyazaki},
  \& {Sugimoto}}]{Oka_2001}
{Oka}, T., {Hasegawa}, T., {Sato}, F., {et~al.} 2001, \apj, 562, 348

\bibitem[{{Papaderos} {et~al.}(1996){Papaderos}, {Loose}, {Thuan}, \&
  {Fricke}}]{Papaderos_1996}
{Papaderos}, P., {Loose}, H.-H., {Thuan}, T.~X., \& {Fricke}, K.~J. 1996,
  \aaps, 120, 207

\bibitem[{{Reines} \& {Deller}(2012)}]{Reines_2012}
{Reines}, A.~E., \& {Deller}, A.~T. 2012, \apjl, 750, L24

\bibitem[{{Reines} {et~al.}(2011){Reines}, {Sivakoff}, {Johnson}, \&
  {Brogan}}]{Reines_2011}
{Reines}, A.~E., {Sivakoff}, G.~R., {Johnson}, K.~E., \& {Brogan}, C.~L. 2011,
  \nat, 470, 66

\bibitem[{{Rice} {et~al.}(2016){Rice}, {Goodman}, {Bergin}, {Beaumont}, \&
  {Dame}}]{Rice_2016}
{Rice}, T.~S., {Goodman}, A.~A., {Bergin}, E.~A., {Beaumont}, C., \& {Dame},
  T.~M. 2016, \apj, 822, 52

\bibitem[{{Rich} {et~al.}(2008){Rich}, {de Blok}, {Cornwell}, {Brinks},
  {Walter}, {Bagetakos}, \& {Kennicutt}}]{Rich_2008}
{Rich}, J.~W., {de Blok}, W.~J.~G., {Cornwell}, T.~J., {et~al.} 2008, \aj, 136,
  2897

\bibitem[{{Rosolowsky}(2005)}]{Rosolowsky_2005}
{Rosolowsky}, E. 2005, \pasp, 117, 1403

\bibitem[{Rosolowsky(2007)}]{Rosolowsky_2007}
Rosolowsky, E. 2007, ApJ, 654, 240

\bibitem[{{Rosolowsky} \& {Leroy}(2006)}]{Rosolowsky_2006}
{Rosolowsky}, E., \& {Leroy}, A. 2006, \pasp, 118, 590

\bibitem[{{Saintonge} {et~al.}(2011){Saintonge}, {Kauffmann}, {Wang}, {Kramer},
  {Tacconi}, {Buchbender}, {Catinella}, {Graci{\'a}-Carpio}, {Cortese},
  {Fabello}, {Fu}, {Genzel}, {Giovanelli}, {Guo}, {Haynes}, {Heckman},
  {Krumholz}, {Lemonias}, {Li}, {Moran}, {Rodriguez-Fernandez}, {Schiminovich},
  {Schuster}, \& {Sievers}}]{Saintonge_2011}
{Saintonge}, A., {Kauffmann}, G., {Wang}, J., {et~al.} 2011, \mnras, 415, 61

\bibitem[{Sandstrom {et~al.}(2013)Sandstrom, Leroy, Walter, Bolatto, Croxall,
  Draine, Wilson, Wolfire, Calzetti, Kennicutt, Aniano, Donovan~Meyer, Usero,
  Bigiel, Brinks, de~Blok, Crocker, Dale, Engelbracht, Galametz, Groves, Hunt,
  Koda, Kreckel, Linz, Meidt, Pellegrini, Rix, Roussel, Schinnerer, Schruba,
  Schuster, Skibba, van~der Laan, Appleton, Armus, Brandl, Gordon, Hinz,
  Krause, Montiel, Sauvage, Schmiedeke, Smith, \& Vigroux}]{Sandstrom_2013}
Sandstrom, K.~M., Leroy, A.~K., Walter, F., {et~al.} 2013, ApJ, 777, 5

\bibitem[{{Santangelo} {et~al.}(2009){Santangelo}, {Testi}, {Gregorini},
  {Leurini}, {Vanzi}, {Walmsley}, \& {Wilner}}]{Santangelo_2009}
{Santangelo}, G., {Testi}, L., {Gregorini}, L., {et~al.} 2009, \aap, 501, 495

\bibitem[{{Sargent} \& {Searle}(1970)}]{Sargent_1970}
{Sargent}, W.~L.~W., \& {Searle}, L. 1970, \apjl, 162, L155

\bibitem[{{Sauvage} {et~al.}(1997){Sauvage}, {Thuan}, \&
  {Lagage}}]{Sauvage_1997}
{Sauvage}, M., {Thuan}, T.~X., \& {Lagage}, P.~O. 1997, \aap, 325, 98

\bibitem[{{Scoville} {et~al.}(1987){Scoville}, {Yun}, {Sanders}, {Clemens}, \&
  {Waller}}]{Scoville_1987}
{Scoville}, N.~Z., {Yun}, M.~S., {Sanders}, D.~B., {Clemens}, D.~P., \&
  {Waller}, W.~H. 1987, \apjs, 63, 821

\bibitem[{{Solomon} {et~al.}(1987){Solomon}, {Rivolo}, {Barrett}, \&
  {Yahil}}]{Solomon_1987}
{Solomon}, P.~M., {Rivolo}, A.~R., {Barrett}, J., \& {Yahil}, A. 1987, \apj,
  319, 730

\bibitem[{{Sun} {et~al.}(2018){Sun}, {Leroy}, {Schruba}, {Rosolowsky},
  {Hughes}, {Kruijssen}, {Meidt}, {Schinnerer}, {Blanc}, {Bigiel}, {Bolatto},
  {Chevance}, {Groves}, {Herrera}, {Hygate}, {Pety}, {Querejeta}, {Usero}, \&
  {Utomo}}]{Sun_2018}
{Sun}, J., {Leroy}, A.~K., {Schruba}, A., {et~al.} 2018, \apj, 860, 172

\bibitem[{{Thuan}(1991)}]{Thuan_1991}
{Thuan}, T.~X. 1991, in Massive Stars in Starbursts, ed. C.~{Leitherer},
  N.~{Walborn}, T.~{Heckman}, \& C.~{Norman} (the Cambridge University Press),
  349

\bibitem[{{Thuan} {et~al.}(1999){Thuan}, {Lipovetsky}, {Martin}, \&
  {Pustilnik}}]{Thuan_1999}
{Thuan}, T.~X., {Lipovetsky}, V.~A., {Martin}, J.-M., \& {Pustilnik}, S.~A.
  1999, \aaps, 139, 1

\bibitem[{{Tully}(1988)}]{Tully_1988}
{Tully}, R.~B. 1988, {Nearby galaxies catalog} (Cambridge and New York,
  Cambridge University Press), 221

\bibitem[{{Turner} {et~al.}(2000){Turner}, {Beck}, \& {Ho}}]{Turner_2000}
{Turner}, J.~L., {Beck}, S.~C., \& {Ho}, P.~T.~P. 2000, \apjl, 532, L109

\bibitem[{{Utomo} {et~al.}(2015){Utomo}, {Blitz}, {Davis}, {Rosolowsky},
  {Bureau}, {Cappellari}, \& {Sarzi}}]{Utomo_2015}
{Utomo}, D., {Blitz}, L., {Davis}, T., {et~al.} 2015, \apj, 803, 16

\bibitem[{{Vacca} \& {Conti}(1992)}]{Vacca_1992}
{Vacca}, W.~D., \& {Conti}, P.~S. 1992, \apj, 401, 543

\bibitem[{{Vacca} {et~al.}(2002){Vacca}, {Johnson}, \& {Conti}}]{Vacca_2002}
{Vacca}, W.~D., {Johnson}, K.~E., \& {Conti}, P.~S. 2002, \aj, 123, 772

\bibitem[{{Vanzi} {et~al.}(2009){Vanzi}, {Combes}, {Rubio}, \&
  {Kunth}}]{Vanzi_2009}
{Vanzi}, L., {Combes}, F., {Rubio}, M., \& {Kunth}, D. 2009, \aap, 496, 677

\bibitem[{{Williams} {et~al.}(1994){Williams}, {de Geus}, \&
  {Blitz}}]{Williams_1994}
{Williams}, J.~P., {de Geus}, E.~J., \& {Blitz}, L. 1994, \apj, 428, 693

\bibitem[{{Williams} \& {McKee}(1997)}]{Williams_1997}
{Williams}, J.~P., \& {McKee}, C.~F. 1997, \apj, 476, 166

\bibitem[{{Wong} {et~al.}(2011){Wong}, {Hughes}, {Ott}, {Muller}, {Pineda},
  {Bernard}, {Chu}, {Fukui}, {Gruendl}, {Henkel}, {Kawamura}, {Klein},
  {Looney}, {Maddison}, {Mizuno}, {Paradis}, {Seale}, \& {Welty}}]{Wong_2011}
{Wong}, T., {Hughes}, A., {Ott}, J., {et~al.} 2011, \apjs, 197, 16

\end{thebibliography}
